\documentclass[longauth]{aa}
\usepackage{graphicx}
\usepackage{setspace}
\usepackage{txfonts}
\usepackage{hyperref}
\usepackage{natbib,twoopt}
\usepackage{threeparttable}

\bibpunct{(}{)}{;}{a}{}{,}
\begin{document}

\title{The search for disks or planetary objects around directly imaged companions: A candidate around DH Tau B. \thanks{Based on observations collected at Paranal Observatory, ESO (Chile)
Program ID: 095.C-0298, 096.C-0241, 097.C-0865, 198.C-0209, and 0104.C-0327(A) and
on observations collected at LBT Observatory. 
%The LBT is an international collaboration among institutions in the United States, Italy and Germany. LBT Corporation partners are: The University of Arizona on behalf of the Arizona university system; Istituto Nazionale di Astrofisica, Italy; LBT Beteiligungsgesellschaft, Germany, representing the Max-Planck Society, the Astrophysical Institute Potsdam, and Heidelberg University; The Ohio State University, and The Research Corporation, on behalf of The University of Notre Dame, University of Minnesota and University of Virginia.
}}

\author{C. Lazzoni \inst{\ref{1} \and \ref{2}}
\and A. Zurlo \inst{\ref{3} \and \ref{4} \and \ref{5}}
\and S. Desidera \inst{\ref{1}}
\and D. Mesa \inst{\ref{1}}
\and C. Fontanive \inst{\ref{5+} \and \ref{1}}
\and M. Bonavita \inst{\ref{1} \and \ref{6}}
\and S. Ertel \inst{\ref{6+} \and \ref{6++}}
\and K. Rice \inst{\ref{6}}
\and A. Vigan \inst{\ref{5}}
\and A. Boccaletti  \inst{\ref{7}}
\and M. Bonnefoy  \inst{\ref{8}}
\and G. Chauvin \inst{\ref{8*} \and \ref{8}}
\and P. Delorme \inst{\ref{8}}
\and R. Gratton \inst{\ref{1}}
\and M. Houllé \inst{\ref{5}}
\and A.L. Maire  \inst{\ref{10} \and \ref{11}}
\and M. Meyer  \inst{\ref{12} \and \ref{9}} 
\and E. Rickman \inst{\ref{13}}
\and E. A. Spalding \inst{\ref{6++} \and \ref{13+}}
\and R. Asensio-Torres \inst{\ref{10}}
\and M. Langlois  \inst{\ref{15} \and \ref{5}} 
\and A. M\"uller  \inst{\ref{10}}
\and J-L. Baudino \inst{\ref{7} \and \ref{16}}
\and J.-L. Beuzit \inst{\ref{8} \and \ref{5}}
\and B. Biller \inst{\ref{10} \and \ref{6}}
\and W. Brandner  \inst{\ref{10}}
\and E. Buenzli \inst{\ref{9}}
\and F. Cantalloube \inst{\ref{10}}
\and A. Cheetham \inst{\ref{10} \and \ref{13}}
\and M. Cudel \inst{\ref{8}}
\and M. Feldt \inst{\ref{10}}
\and R. Galicher  \inst{\ref{7}}
\and M. Janson \inst{\ref{10} \and \ref{17}} 
\and J. Hagelberg \inst{\ref{13}}
\and T. Henning \inst{\ref{10}}
\and M. Kasper \inst{\ref{8} \and \ref{18}} 
\and M. Keppler \inst{\ref{10}}
\and A.-M. Lagrange \inst{\ref{8}}
\and J. Lannier \inst{\ref{8}}
\and H. LeCoroller \inst{\ref{5}}
\and D. Mouillet \inst{\ref{8}}
\and S. Peretti \inst{\ref{13}}
\and C. Perrot  \inst{\ref{7}}
\and G. Salter  \inst{\ref{5}}
\and M. Samland  \inst{\ref{10}}
\and T. Schmidt  \inst{\ref{7}}
\and E. Sissa  \inst{\ref{1}}
\and F. Wildi  \inst{\ref{13}}
}

\institute{
        INAF -- Osservatorio Astronomico di Padova, Vicolo dell'Osservatorio 5, I-35122, Padova, Italy \label{1}
        \and Dipartimento di Fisica a Astronomia "G. Galilei", Universita' di Padova, Via Marzolo, 8, 35121 Padova, Italy \label{2}
        \and N\'ucleo de Astronom\'ia, Facultad de Ingenier\'ia y Ciencias, Universidad Diego Portales, Av. Ejercito 441, Santiago, Chile \label{3}
        \and Escuela de Ingenier\'ia Industrial, Facultad de Ingenier\'ia y Ciencias, Universidad Diego Portales, Av. Ejercito 441, Santiago, Chile \label{4}
        \and Aix Marseille Univ, CNRS, LAM, Laboratoire d'Astrophysique de Marseille, Marseille, France \label{5}
        \and Center for Space and Habitability, University of Bern, 3012 Bern, Switzerland \label{5+}
                \and Institute for Astronomy, The University of Edinburgh, Royal Observatory, Blackford Hill, Edinburgh, EH9 3HJ, U.K. \label{6}
        \and Large Binocular Telescope Observatory, 933 North Cherry Avenue, Tucson, AZ 85721, USA \label{6+}
        \and Steward Observatory, Department of Astronomy, University of Arizona, 993 N. Cherry Ave, Tucson, AZ, 85721, USA \label{6++}
        \and LESIA, Observatoire de Paris, PSL Research University, CNRS, Sorbonne Universités, UPMC Univ. Paris 06, Univ. Paris Diderot, Sorbonne Paris Cité \label{7}
        \and Universit\'e Grenoble Alpes, IPAG, 38000 Grenoble, France \label{8}
                    \and Department of Astronomy, University of Chile, Casilla 36-D, Santiago, Chile \label{8*}
        \and Max-Planck Institute for Astronomy, K\"onigstuhl 17, 69117 Heidelberg, Germany \label{10}
        \and STAR Institute, Universit\'{e} de Li\`{e}ge, All\'{e}e du Six Ao\^{u}t 19c, 4000, Li\`{e}ge, Belgium \label{11}
        \and Department of Astronomy, University of Michigan, 1085 S. University, Ann Arbor, MI 48109 \label{12}
        \and Institute for Astronomy, ETH Zurich, Wolfgang-Pauli Strasse 27, 8093 Zurich, Switzerland \label{9}
            \and Observatoire Astronomique de l'Universit\'e de Gen\`eve, Chemin des Maillettes 51, 1290 Sauverny, Switzerland \label{13}
            \and Center for Astronomical Adaptive Optics, University of Arizona, Tucson, AZ 85721 \label{13+}
        \and CRAL, UMR 5574, CNRS, Universit\'{e} Lyon 1, 9 avenue Charles Andr\'{e}, 69561 Saint Genis Laval Cedex, France \label{15}
        \and Department of Physics, University of Oxford, Oxford OX1 3PU, UK \label{16}
        \and Department of Astronomy, Stockholm University, AlbaNova University Center, 106 91 Stockholm, Sweden \label{17}
        \and European Southern Observatory, Karl Schwarzschild St, 2, 85748 Garching, Germany \label{18}
       }

\abstract {In recent decades, thousands of substellar companions have been discovered with both indirect and direct methods of detection. While the majority of the sample is populated by objects discovered using radial velocity and transit techniques, an increasing number have been directly imaged. These planets and brown dwarfs are extraordinary sources of information that help in rounding out our understanding of planetary systems.} {In this paper, we focus our attention on substellar companions detected with the latter technique, with the primary goal of investigating their close surroundings and looking for additional companions and satellites, as well as disks and rings. Any such discovery would shed light on many unresolved questions, particularly  with regard to their possible formation mechanisms.} {To reveal bound features of directly imaged companions, whether for point-like or extended sources, we need to suppress the contribution from the source itself. Therefore, we developed a method based on the negative fake companion (NEGFC) technique that first estimates the position in the field of view (FoV) and the flux of the imaged companion with high precision, then subtracts a rescaled model point spread function (PSF) from the imaged companion, using either an image of the central star or another PSF in the FoV. Next it performs techniques, such as angular differential imaging (ADI), to further remove  quasi-static patterns of the star (i.e., speckle contaminants) that affect the residuals of close-in companions.} {After testing our tools on simulated companions and disks and on systems that were chosen ad hoc,  we applied the method to the sample of substellar objects observed with SPHERE during the SHINE GTO survey. Among the 27 planets and brown dwarfs we analyzed, most objects did not show remarkable features, which was as expected, with the possible exception of a point source close to DH Tau B. This candidate companion was detected in four different SPHERE observations, with an estimated mass of $\sim 1$ M\textsubscript{Jup}, and a mass ratio with respect to the brown dwarf of $1/10$. This binary system, if confirmed, would be the first of its kind, opening up interesting questions for the formation mechanism, evolution, and frequency of such pairs. In order to address the latter, the residuals and contrasts reached for 25 companions in the sample of substellar objects observed with SPHERE were derived. If the DH Tau Bb companion is real, the binary fraction obtained is $\sim 7\%$, which is in good agreement with the results obtained for field brown dwarfs.  } {While there may currently be many limitations affecting the exploration of bound features to directly imaged exoplanets and brown dwarfs, next-generation instruments from the ground and space (i.e., JWST, ELT, and LUVOIR) will be able to image fainter objects and, thus, drive the application of this technique in upcoming searches for exo-moons and circumplanetary disks.}


\titlerunning{Disks or satellites around DI companions}
\authorrunning{Lazzoni et al.}
\maketitle

\section{Introduction}
To improve our knowledge and comprehension of how planets form and evolve in their environments,  their close neighborhoods must be investigated. In our own Solar System, planets are often surrounded by satellites and disk-like features, such as dusty rings \citep{Alibert}. Moreover, these elements seem to increase in number and extensions with the mass of the central planet. Although there has been a tentative detection of a Saturn-like moon around Kepler-1625 b \cite{Teachey} and a report of a disk or ring-like system candidate  orbiting the pre-main-sequence star 1SWASP J140747.93-394542.6 from a complex photometric dimming event \citep{mamajek2012}, there has been no exomoon, or exoring, confirmed yet.
Another case that is still under debate is that of Fomalhaut b \citep{Lawler}, in which the spectral energy distribution was interpreted as the presence of a cloud of dust and planetesimals around a low-mass planet, indicating the presence of an extended source, even if more recent studies have suggested that Fomalhaut b is not a planet but the result of a collision between two large planetesimals \citep{Gaspar}.

With the current instrumentation and telescopes, the hunt for features comparable to those observed in the Solar System is considerably challenging for any detection technique. However, we would expect the mass of a disk surrounding a substellar object to scale with the mass of the companion itself \citep{Magni}; satellites and ring systems should also follow such trends \citep{Ward}. Thus, the ideal targets for investigation for bound features would be massive companions spanning the regime from a few Jupiter masses to brown dwarfs. The detectability is further improved for young systems in which planets are still in the thermal emission regime and circumplanetary disks are more likely to be present. Direct imaging instruments such as GPI \citep{Macintosh} and SPHERE \citep{Beuzit} might be used at the extremity of their potential in the search for features around directly imaged planets and brown dwarfs. 

Additional advantages arise when considering giant planets and brown dwarfs at wide separations. Indeed,
the gravitational influence of a substellar object, as measured by the radius of the Hill Sphere, increases with the mass and the separation from the host star. Then, not only can we expect to find bound features (disks, rings, satellites, etc.) placed at larger distances with respect to such massive and wide companions but also the probability of interaction and pair formation of planets and brown dwarfs is enhanced. 
For example, the Hill radius of a Jupiter-mass planet at 1 au around a Sun-like star is roughly 0.07 au, whereas, the same planet at 50 au has an Hill radius of more than 3 au. A 50 M\textsubscript{Jup} brown dwarf at the same separations has an Hill radius of 0.25 au and 12.57 au, respectively. Since typical separations of objects detected with the direct imaging technique are in tens of au, we may find further companions inside their Hill sphere up to few au and this would allow us to disentangle the point spread functions (PSFs) of the two objects with 8-meter class telescopes equipped with xAO. Therefore, even if the terrestrial regime is almost inaccessible for this generation of instruments, we can still constrain upper limits for the mass of the satellites, look for binary planets, and investigate the presence of extended structures such as disks and rings.

Pairs of brown dwarfs in wide orbits around stars have been discovered in several studies, such as  $\epsilon$ Indi B \citep{King}, GJ 569 B \citep{Femenia}, GJ 417 B \citep{Kirkpatrick}, HD130948 B \citep{Potter} and AB Dor Ca/Cb \citep[via interferometry, ][]{Climent}. 
This kind of configuration might extend to massive giant planets and brown dwarfs close to deuterium burning limits. 
Isolated brown dwarfs or brown dwarf-giant planet pairs have been already reported as well \citep[e.g., Ophiucus 1622 2405 and 2MASS J1207334393254, ][]{Luhman,Chauvin3}.

On the disk side, there are a number of confirmed cases of substellar companions down to planetary masses showing significant H$\alpha$ or Paschen $\beta$ emission. These are considered robust evidence of on-going accretion, as in the case of PDS 70 b \citep{Wagner, Christiaens} and c \citep{haffert2019}, GQ Lup B \citep{Seifahrt}, DH Tau B \citep{Bonnefoy3}, SR 12 C \citep{Santamaria}, and USCO1610 B \citep{Petrus}, although in certain other cases the observed emission is most likely due to chromospheric activity  \citep{Barcucci,Petrus}. Some of these candidate circumplanetary disks were also confirmed in the submillimeter with ALMA, \citep[PDS70 c, ][]{Isella}, while others, such as those surrounding GQ Lup B \citep{MacGregor2} and DH Tau B \citep{Long,Wolff, Perez}, escaped detection at this wavelength. 

Disk or ring-like features may form at later phases, as a result of collisions that could happen either within a satellite system or between a planet or satellites and an incoming body \citep{Wyatt1}. The latter case may happen more frequently for stars surrounded by debris disks that scatter large bodies inwards and outwards, increasing the chances of collisions with planets or satellites \citep{Hyodo}.

In this paper, we present routines based on the fake negative companion technique to suppress the contribution of directly imaged sources and investigate the residuals. These tools were systematically applied to every suitable substellar object, detected during the SHINE/SPHERE survey \citep{Chauvin2} to unveil possible extended structures and to constrain their multiplicity. Since typical contrasts obtained with SPHERE are on the order of $10^{-5}-10^{-6}$, for the youngest systems, the instrument can probe down to a few Jupiter masses. Thus, as mentioned before, with the current instrumentation, we are not sensitive to Solar System-like rocky satellites but only to gas giant companions and to bright and massive accretion disks. In this framework, we present the first indication of a Jupiter-like candidate that orbits a low-mass brown dwarfs bound to DH Tau.

In this paper, we present a description of the method in Sect. 2 . In Sect. 3, we investigate the entire SPHERE sample and in Sect. 4, we present the DH Tau Bb candidate companion. In Sect. 5, we show the detection limits and the statistics obtained for the sample of substellar objects observed with SPHERE. In Sect. 6, we present our conclusions. Finally, in App. A we present results obtained by applying our tools to synthetic data and \textbf{to} systems \textbf{chosen ad hoc}. In App. B, we describe the method used for the reconstruction of the PSF of the star behind the coronagraph and in App. C, we describe the reduction of the LMIRCam/LBTI data for DH Tau B.

\section{Methods}

\subsection{The NEGFC technique}
In order to unveil the neighborhood of known companions detected with direct imaging techniques, we need to carefully subtract the contribution of the companion itself. We applied the procedure known as the negative fake companion technique \citep[NEGFC,][]{Lagrange3, Marois4} which, in short, consists of forward modeling a detected source by injecting a negative model of the instrument PSF into the raw data and minimizing the residuals in the final image after post-processing by adjusting the PSF position and flux.

As model PSF, we used, in general, the off-axis image of the central star, which resemble the shape of the PSF of the companion of interest. The frame containing the model should have dimensions that are sufficient to include several times the full width half maximum (FWHM) of the PSF. This is needed in order to reduce artifacts in the close vicinity of the companion when subtracting the model. In SPHERE observations, for example, the PSF template for IRDIS (see Section 3) is an image of $64\times64$ pixels (roughly 16 times the FWHM in the H-band and 12 times in the K-band) of the central star, which is provided by the SPHERE Data Center \citep{Delorme2}. Moreover, all our tools rely on the angular differential imaging technique \citep[ADI,][]{Marois5} but they could also be used, in principle, with different post-processing techniques (for example, a median of the frames).

We thus implemented the NEGFC in Python, starting from the already released VIP code \citep{Gonzalez} and optimizing some of its functions to look for both bound extended features and point-like sources around directly imaged companions. For the faintest objects and companions located in the innermost regions dominated by speckle contaminants, we assumed priors for the $(r,\theta, F)$ of the planet, subtracted a model of the latter in each raw frame, and calculated the $\chi^2$ on a circular aperture centered on the companion in the post processed image obtained with specific techniques, such as classical ADI \citep[cADI,][]{Marois5} and the principal component analysis \citep[PCA,][]{Soummer2, Amara}. The $\chi^2$ is defined as a sum over the $N$ positive values divided by $N-3,$ if the function of merit is \textit{sum}, or as the standard deviation over the values different from zero, if the function of merit is \textit{stddev}. The $(r,\theta, F)$ parameters are then optimized to minimize the $\chi^2$. Here, we call this procedure the \textit{single\_fullcube} routine.

\subsection{Evolutions of NEGFC}

 Since the \textit{single\_fullcube} routine is not able to take into account the frame-by-frame variations of the flux due to changing atmospheric conditions, we propose a first evolution of NEGFC for companions visible in each raw frame ($S/N\gtrsim 30$). The \textit{single\_framebyframe} routine determines $(r,\theta,F)$ for each frame of the unprocessed dataset.
The \textit{single\_fullcube} routine is then used to suppress the contribution of the faintest or closest companion, whereas the \textit{single\_framebyframe} tool is applied at the brightest or farthest point sources that are detectable in each raw frame and that are clearly distinguishable from speckle contaminants. 

For ground-based observations, the shape of PSFs over time is not constant. Indeed, if parameters such as seeing or wind speed change significantly during the observation, the PSF of the companion might be quite different from the first to the last frame of the cube. Also, in order to reach deeper contrasts,  the coronagraphic observations are very often the best choice for exoplanets search. With this kind of setting, the star is hidden behind the coronagraph and cannot be used as a model. Thus, we need an image of the off-axis PSF of the star that might nonetheless differ significantly from the PSF of the companion since the two are not taken simultaneously.  When subtracting two PSFs with different shapes induced by surrounding conditions, spurious structures may emerge in the residuals. For coronagraphic observations, then, the process of estimating the position and flux of directly imaged companions and the retrieval of features in their residuals works more efficiently for datasets taken in good and stable weather conditions.

When, instead, the observations are taken without the coronagraph, we can use the the star itself as a model PSF for each of the frames in the cube. This second evolution of the NEGFC, which uses more than one model PSF,  is called the \textit{multiple\_framebyframe} routine here. This tool helps to avoid eventual artifacts that may emerge from changes in the conditions during the observation. Indeed, the model PSF and the PSF of the companion are taken at the same time and, thus, they have the same shape if their separation is not too large (otherwise, anisoplanatism effects may elongate the PSF and decrease the Strehl ratio). On the other hand, with these kind of settings, the processed images reach poor contrasts compared to a coronagraphic sequence so that in order to distinguish structures around the companion, or even the companion itself, we need to concentrate on brighter planets or brown dwarfs. 

One further possibility to use the \textit{multiple\_framebyframe} routine concerns datasets where there are more companions or known background stars in the FoV of the instrument. To be useful as models, such objects should be located in regions of low speckle contaminant noise and with a similar radial separation as the one of the companion under investigation. In any case, the companion used to generate the model PSFs should be very bright with a signal to noise ratio (S/N) on the order of 50 or greater. If this is the case, the peak is expected to be well above the background and detectable in each raw frame. Therefore, suitable point sources in the FoV to be converted into model PSFs are quite rare. We show in Section 3.3 an example of the gain in detection limits and improvement of the residuals for the HD 1160 system. This was the only case with a point-like source in the FoV that was bright enough to apply this routine. 

It is worth mentioning the four replicas of the PSF of the central star that are often used to measure the accurate position of the star behind the coronagraph. These satellite spots would be, in principle, perfect candidates of model PSFs present in each frame of the cube \citep{Marois3, Langlois}, especially for monochromatic observations for which there are no radial elongations. Unfortunately, they are placed very close to the outer working angle of the adaptive optics (AO), where speckles dominate the background. Indeed, based on the preliminary test run on real data,  the results demonstrate that the subtraction of the companion using satellite spots with the \textit{multiple\_framebyframe} is worse than the one obtained with the single-PSF routines.

The output of each routine consists of: 1) an image of the companion obtained from a post processed image (median, cADI, or PCA+ADI) of the original cube; 2) an image of the model obtained from the post-processing of an empty cube where the model PSF is added in each frame at the position and with the flux obtained for the companion; 3) an image of the residuals around the companion, obtained by processing (either with a median, cADI, or PCA+ADI) the scientific cube \textbf{from which} the companion was subtracted.

An additional output of each routine is the contrast curve centered on the planet or the brown dwarf. The detection limits are obtained by re-centering the post-processed frame from which the companion was subtracted at the position of the companion itself and calculating the standard deviation in concentric annuli with a width of one FWHM each, starting from a separation of 1/2 FWHM from the center. The limits for the detection are set at the $5 \sigma $ level and they include the small sample statistics following the discussion presented in \cite{Mawet} and the algorithm throughput effects \citep{2018AJ....155...19J}.

\subsection{Tests on synthetic and real companions.}

As a first step, we tested our routines on synthetic companions. A more detailed discussion is presented in Appendix A.1. The most evident result is the gain in contrast close to the PSF of the injected companion once the latter is subtracted. Furthermore, we simulated a disk surrounding the synthetic companion with different inclinations and fixed contrast and separation. The inclination of the system is, indeed, a fundamental parameter for the detection of extended features, together with the luminosity and the separation from the companion (see Appendix A.2). As a result, we found that disk-like sources would also, in principle, be detectable using the method presented in this paper.

We also tested our tools on real companions choosing two ad hoc datasets. The first system we analyzed was the crowded FoV of HIP 87836, with many suitable PSFs that show signs of binarity (see Appendix A.3). The second system was, instead, chosen to prove the efficiency of the method to detect extended features. We analyzed the HST observation of the additional stellar companion that was recently discovered in orbit around GQ Lup at wide separations. GQ Lup C \citep{Alcala} showed evidence of the presence of a disk based on the infrared excess in its spectral energy distribution. We then applied the \textit{single\_framebyframe} routine to resolve the disk in reflected light. In each of the nine frames (one for each wavelength), we subtracted a model obtained by a nearby star from the PSF of the companion. The disk around GQ Lup C was easily detected with an estimated width of 88 mas and an inclination of $44^{\circ}$ \citep{Lazzoni}.

\section{SPHERE data}
\subsection{The SPHERE instrument}

The components of SPHERE include an extreme adaptive optics system, SAXO \citep{Fusco, Petit}, using 41 x 41 actuators, pupil stabilization, and differential tip-tilt control. The instrument has several coronagraphic devices for stellar diffraction suppression, including apodized pupil Lyot coronagraphs \citep{Carbillet} and achromatic four-quadrant phase masks \citep{Boccaletti}. 

Among the science subsystems available, we used only the Infra-Red Dual-band Imager and Spectrograph \citep[IRDIS,][]{Dohlen} observations, since IRDIS has a wider field of view and most of our targets are placed at few arcseconds. The stars in our sample were observed in IRDIFS mode, with IRDIS in dual-band imaging mode \citep[DBI;][]{Vigan1}, using the narrow-band H2H3 filters ($\lambda=1.593, 1.667\mu m$), or in IRDIFS\_EXT mode, using the K1K2 filters ($\lambda=2.110 , 2.251\mu m$) of the instrument. For one of the observations of HR 2562, the broad band H filter ($\lambda= 1.625 \mu m$) was used.

The IRDIS data were pre-reduced at the SPHERE data center \citep{Delorme2}, hosted by OSUG/IPAG in Grenoble, using the SPHERE Data Reduction Handling (DRH) pipeline \citep{Pavlov} and the dedicated Specal data reduction software \citep{Galicher}

\begin{table*}
 \caption{Observing log. The asterisks indicate the datasets used to derive the flux of the companions and the contrast curves  when multiple observations were available for a single system.}
\label{tabu1}   
\centering
\begin{threeparttable}
\begin{tabular}{c c c c c c}
\hline\hline
Name & Obs Date & Observing Mode &Tot rotation angle & Seeing & Reference \\
\hline
\\  
51 Eri      &2017-09-27&   IRDIFS\_EXT&  $46.4^{\circ}$ &    $0.49''$     &   a     \\[0.75ex]  
AB Pic     & 2015-02-06&IRDIFS&  $28.3^{\circ}$ &  $1.10''$ & b \\[0.75ex]  
$\beta$ Pic & 2015-02-05&IRDIFS&$76.6 ^{\circ}$&  $0.77''$ & c \\[0.75ex] 
CT Cha   & 2017-03-19&IRDIFS& $25.0^{\circ}$&$ 0.77'' $ & d \\[0.75ex]  
DH Tau   & 2015-10-26&IRDIFS&$20.0 ^{\circ}$& $1.27''$ & \\[0.75ex]  
               & *2015-12-19&IRDIFS&$24.5 ^{\circ}$&$ 0.63''  $ &\\[0.75ex]  
               &2018-12-16&IRDIFS&$33.7^{\circ}$ & $0.48''$& \\[0.75ex]
               &2019-09-19&LMIRCam&$37.0^{\circ}$ & $0.65''$& \\[0.75ex]
               &2020-01-05&IRDIFS\_EXT&$17.4^{\circ}$ & $0.55''$& \\[0.75ex]
$\eta$ Tel  &2015-05-04 &IRDIFS& $41.5^{\circ}$ &$ 1.11'' $&b \\[0.75ex]
GJ 504   &*2015-06-03 &IRDIFS& $28.9^{\circ}$ &$ 1.46'' $ &e\\[0.75ex] 
               &2016-03-29 &IRDIFS& $27.5^{\circ}$ & $ 0.98''$ &e \\[0.75ex] 
GQ Lup   & 2015-05-04&IRDIFS& $25.1^{\circ}$ & $1.05''$ & \\[0.75ex] 
               & *2016-06-26&IRDIFS& $90.9 ^{\circ}$& $0.61'' $ &\\[0.75ex] 
HD1160    &2015-07-01 &IRDIFS& $30.8 ^{\circ}$& $0.53''$  \\[0.75ex]
HD4747   & 2016-12-11&IRDIFS\_EXT& $7.1^{\circ}$ &$2.11'' $ & f  \\[0.75ex] 
HD19467   &2017-11-03 &IRDIFS\_EXT& $63.3 ^{\circ}$& $ 0.52'' $ &g\\[0.75ex]
HD95086  & 2018-01-05&IRDIFS\_EXT&$ 41.0^{\circ}$&  $0.30''$& \\[0.75ex] 
HIP64892   & 2016-04-01& IRDIFS&$44.7^{\circ}$ &  $0.83'' $ & h\\[0.75ex] 
                   & *2017-02-08& IRDIFS\_EXT&$54.8 ^{\circ}$&  $0.80''$ & h \\[0.75ex] 
HIP65426   & 2016-06-26&IRDIFS&$42.2^{\circ}$ &  $ 0.73''$ & i\\[0.75ex] 
                   & *2017-02-06&IRDIFS&$44.3 ^{\circ}$&  $0.60''$ & i \\[0.75ex] 
                   & 2018-05-12&IRDIFS\_EXT&$ 31.6^{\circ}$&  $0.81''$ & j \\[0.75ex] 
HIP78530   &2015-05-04 & IRDIFS&$36.3^{\circ}$ & $ 0.66''$ & b\\[0.75ex]
HIP107412   &2016-09-16 & IRDIFS\_EXT&$75.9 ^{\circ}$& $0.68''$ & k  \\[0.75ex]
HR2562   &*2017-02-06 &BROAD BAND H&$34.9^{\circ}$ & $ 0.54'' $ & l\\[0.75ex]
              &2017-09-28 & IRDIFS\_EXT&$27.6^{\circ}$ & $0.78'' $& m\\[0.75ex]
HR3549  &2015-12-19 & IRDIFS& $32.5^{\circ}$& $ 0.91''$ & n\\[0.75ex]
HR8799 &2015-07-04 & IRDIFS\_EXT&$19.1^{\circ}$ & 1.05''&o\\[0.75ex]
PDS 70& 2018-02-04&IRDIFS\_EXT&$ 95.7^{\circ}$&  $0.41''$ &p \\[0.75ex] 
PZ Tel   &*2015-05-06 & IRDIFS& $11.4^{\circ}$& $1.24'' $&b \\[0.75ex] 
             &2015-05-31 & IRDIFS& $9.3^{\circ}$& $ 1.30''$ & b\\[0.75ex] 
TYC 7084-794-1& 2015-11-29& IRDIFS& $38.9^{\circ}$& $1.01'' $&b \\[0.75ex]  
                     & *2016-01-19& IRDIFS\_EXT&$70.6^{\circ}$ &  $1.17'' $& b\\[0.75ex]  
TYC 8047- 232-1 & *2015-09-25&IRDIFS&$81.5^{\circ}$ &  $0.99'' $&b\\[0.75ex]  
                          &2016-01-17 &IRDIFS&$14.3^{\circ}$ &$ 1.58''$ &b \\[0.75ex]  
\\[0.75ex]  
HIP87836 & 2018-05-09&IRDIFS&$50.4^{\circ}$ &  $ $\\[0.75ex]                            
\hline
\end{tabular}

\tablefoot{References: \textbf{a} \cite{Maire2}; \textbf{b} Langlois et al. in prep; \textbf{c} \cite{Lagrange4}; \textbf{d} Schmidt et al. in prep; \textbf{e} \cite{Bonnefoy2}; \textbf{f} \cite{Peretti}; \textbf{g} Maire et al. in prep; \textbf{h} \cite{Cheetham}; \textbf{i} \cite{Chauvin}; \textbf{j} \cite{Cheetham2}; \textbf{k} \cite{Delorme}; \textbf{l} \cite{Mesa2}; \textbf{m} \citep{Maire4}; \textbf{n} \cite{Mesa3}; \textbf{o} Zurlo et al. in prep; \textbf{p} \cite{Muller} }

\end{threeparttable}
\end{table*}

\begin{table*}
\caption{Projected distance and contrast values were obtained from the observations presented in the previous Table \citep{Maire3} (for systems with multiple observations, parameters were derived from datasets with asterisks); masses were obtained using BT SETTL models and for 51 Eri b and HD 95086 b, the AMES COND models; distances were taken from the GAIA catalogue \citep{GAIA,GAIA1}; references for the ages are listed in the notes.}
\label{tabu2}   
\centering
\begin{threeparttable}
\begin{tabular}{c c c c c c c c}
\hline\hline
Name & Separation & Contrast & Age & Mass& Distance & Hill Radius & Age References \\
           &    &&(Myrs)&($M_{jup}$)&(pc)&(au)&\\
\hline
\\  
51 Eri b  &  $0.45 ''$  &    $5.8\times10^{-6}$  &  $24\pm 5$  &  $3.4_{-1.4}^{+0.4}$  &  $29.8$  &1.2& a \\[0.75ex]    %beta pic MG
               
AB Pic B  &  $ 5.40''$  &    $6.4\times10^{-4}$  &  $45_{-10}^{+5}$  &  $14.0_{-0.4}^{+0.3}$  & $ 50.1 $ &40.8 & b\\[0.75ex]              %carina MG
  
$\beta$ Pic b &  $ 0.33''$  &    $1.0\times10^{-4}$  &  $24\pm 5$  &  $11.8_{-0.6}^{+0.7}$  &  $19.7$ &1.1& a\\[0.75ex]        %beta pic MG
    
CT Cha B  &  $ 2.68''$  &  $1.7\times10^{-3}$  &  $1.41_{-0.30}^{+0.38}$  &  $15_{-1}^{+2}$  &$ 191.8$ & 78.5 & c \\[0.75ex]  %Feiden 2016

DH Tau B  &  $ 2.35''$  &    $4.1\times10^{-3}$   &  $1.4\pm0.1$   &  $10.6_{-0.3}^{+0.4}$  &  $135.3$ & 66.4 & c \\[0.75ex]   %Feiden 2016

$\eta$ Tel B  &  $ 4.21''$  &   $1.5\times10^{-3}$  &  $24\pm5$  &  $47_{-6}^{+5}$  &  $47.4$& 34.4 & a \\[0.75ex]     %beta pic MG

GJ504 B  &  $2.49''$  &    $1.3\times10^{-6}$  &  $4000\pm1800$  &  $25_{-7}^{+5}$  & $ 17.5$ & 7.6 & d  \\[0.75ex] %Bonnefoy 2018 

GQ Lup B  &  $0.70''$  &    $3.1\times10^{-3}$  &  2-5   &  $27.6_{-2.6}^{+0.8}$  &  $151.8 $ & 24.7 & e \\[0.75ex]   %Donati 2012

HD1160 B   &  $0.78''$  &   $8.9\times10^{-4}$  &  $50_{-40}^{+50}$  &  $58_{-38}^{+23}$  & $ 125.9 $& 42.8 &f \\[0.75ex] %Maire 2016 (Curtis 2019 consistent with 120)

HD1160 C   &  $5.15''$  &   $4.4\times10^{-3}$  &  $50_{-40}^{+50}$  &  $175_{-100}^{+68}$  & $ 125.9 $ & 409.9 &f\\[0.75ex]   %Maire 2016 (Curtis 2019 consistent with 120)

HD4747 B  &  $0.59''$  &    $6.5\times10^{-4}$  &  $2300\pm1400$  &  $72_{-16}^{+3}$  &  $18.8$& 1.0 &g  \\[0.75ex]   %Peretti 2018

HD19467 B  &  $1.63''$    &  $2.4\times10^{-5}$  &  $8000_{-1000}^{+2000}$  &  $70.3_{-3.7}^{+0.9}$  &  $32.0$ & 14.9 &h \\[0.75ex]                %Maire et al. in prep

HD95086 b  &  $ 0.62''$  &   $9.1\times10^{-6}$  &  $26_{-14}^{+24}$  &  $4_{-2}^{+2}$  &  $86.4$ & 3.9 &i \\[0.75ex]   %Desidera in prep 
          
HIP64892B  &  $1.27''$  &    $1.3\times10^{-3}$  &  $16_{-7}^{+15}$  &  $41_{-17}^{+23}$  & $ 125.2 $ & 28.2 &j \\[0.75ex]  %Cheetham 2018 

HIP65426 b  &  $ 0.83''$  &    $3.4\times10^{-5}$  &  $14\pm4$  &  $7.1_{-0.3}^{+0.4}$  & $ 109.2 $& 12.1 &k \\[0.75ex]      %Chauvin 2017   

%HIP74865 B  &  $0.13''$  &   $3.3\times10^{-3}$  &  $17_{-2}^{+3}$  &  $42.25_{-7.21}^{+5.54}$   &  $123.5$ & 3.4 &l \\[0.75ex]   %UCL MG  \textbf{l} UCL Moving Group;

HIP78530 B  &  $4.18''$    &  $7.1\times10^{-4}$  &  $11_{-4}^{+12}$  &  $20_{-1}^{+18}$  & $ 137.3 $ &113.1 &l \\[0.75ex]    %MG??? 

HIP107412 B  &  $0.27''$   &  $1.2\times10^{-4}$  &  50-700  &  $31_{-21}^{+11}$   & $ 40.8 $ &2.2 &m\\[0.75ex]     %Delorme 2017

HR2562 B  &  $0.64''$  &    $2.3\times10^{-5}$  & 200-750  &  $26_{-2}^{+7}$  & $ 34.0$ & 4.3 &n \\[0.75ex]         %Mesa 2018

HR3549 B  &  $0.85''$  &   $2.1\times10^{-4}$  &  100-150  &  $48_{-5}^{+3}$  & $ 95.4$ & 15.2&o \\[0.75ex]         %Mesa 2016

HR8799 b  &  $1.72''$  &   $2.8\times10^{-5}$  &  $42_{-7}^{+8}$  &  $6.77_{-0.03}^{+0.74}$  &  $41.3 $ & 7.1 & p  \\[0.75ex]         %columba MG

HR8799 c  &  $0.95''$  &    $6.9\times10^{-5} $  &  $42_{-7}^{+8}$  &  $9.3_{-0.1}^{+0.3}$  &$  41.3 $ & 4.9 &p \\[0.75ex]        %columba MG

HR8799 d  &  $0.66''$  &    $7.2\times10^{-5}$  &  $42_{-7}^{+8}$  & $9_{-0.2}^{+0.2}$  &  $41.3$  & 2.6 &p \\[0.75ex]        %columba MG

HR8799 e  &  $ 0.39''$  &    $7.6\times10^{-5}$  &  $42_{-7}^{+8}$  &  $9.6_{-0.2}^{+0.2}$  & $ 41.3 $ & 1.8 &p \\[0.75ex]        %columba MG

PDS 70 b  &  $ 0.19''$  &  $5.5\times10^{-4}$  &  $5.4\pm1.0$  &  $5.6_{-0.4}^{+0.6}$  &  $113.4$ & 2.9 &q \\[0.75ex]     %Keppler 2018
       
PZ Tel B  &  $0.50''$  &    $3.0\times10^{-3}$  &  $24\pm5$  &  $30_{-10}^{+6}$  & $ 47.1 $ & 1.4 &a \\[0.75ex]    %beta pic MG

TYC 7084-794-1 B  &  $2.99''$    &  $4.0\times10^{-3}$  &  $140\pm40$  &  $32_{-4}^{+5}$  & $ 22.4 $ & 19.8& r \\[0.75ex]            %AB Dor MG

TYC 8047-232-1 B  &  $3.21''$   &  $8.0\times10^{-4}$  &  $42_{-7}^{+8}$  &  $13.8_{-0.3}^{+0.4}$  &  $86.3$ & 47.0 &p \\[0.75ex]       %columba MG

\hline

\end{tabular}
\tablefoot{\textbf{a} $\beta$ Pic Moving Group;  \textbf{b} Carina Moving Group;  \textbf{c} \cite{Feiden};  \textbf{d} \cite{Bonnefoy2};  \textbf{e} \cite{Donati};  \textbf{f} \cite{Nielsen};  \textbf{g} \cite{Peretti};  \textbf{h} Maire et al. in prep.;  \textbf{i} Desidera et al. in prep.;  \textbf{j} \cite{Cheetham};  \textbf{k} \cite{Chauvin};  \textbf{l} Upper ScoCen Moving Group; \textbf{m} \cite{Delorme};  \textbf{n} \cite{Mesa2};  \textbf{o} \cite{Mesa3}; \textbf{p} Columba Moving Group;  \textbf{q} \cite{Keppler};  \textbf{r} AB Doradus Moving Group.}

\end{threeparttable}

\end{table*}

\subsection{The SPHERE sample}
The methods and routines described and tested in the previous Section and in the Appendix have as their primary objective the discovery of (sub)planetary satellites and disks around companions detected with the direct imaging technique. For this reason, we analyzed residuals around substellar companions detected with SPHERE/VLT during the SpHere INfrared survey for Exoplanets (SHINE, Desidera et. al, in prep.) guaranteed time of observation (GTO) for 27 young systems hosting one or more companions, with masses spanning from Jupiter to brown dwarfs range.  Among the sample we discarded TWA 5 B and HD 284149 B because the central stars are close visual binaries \citep{Macintosh2, Bonavita1}, thus the model PSF obtained from the latter is not suitable for our purposes, along with ROXs42 B and GSC 6214-210 B due to their poor or short datasets. 

In Table \ref{tabu1}, we provide some brief information on the observations of the 23 systems in the sample and in Table \ref{tabu2}, we list the characteristics of the 27 objects in the analysis. Several of the datasets have already been published in papers by the SHINE team (references in Table \ref{tabu1}) and others will be presented in detail in forthcoming papers that are currently in preparation.
The ages were retrieved either from the literature or from those adopted in SPHERE's statistical paper \citep[Langlois et al., in prep: Desidera et al., in prep.; for a preliminary report see][]{Meyer} for the objects observed in the first half of the survey.

\subsection{Results}
\begin{figure} [h!]
\flushleft
\includegraphics[scale=0.35]{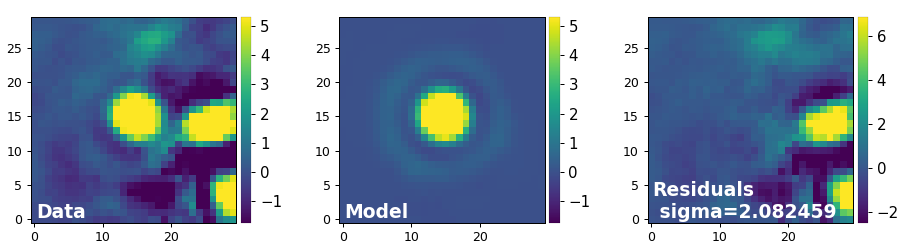}
\caption{cADI image of HIP107412, the model, and the residuals}
\label{HIP107412}
\end{figure}

Among the objects analyzed in the sample, we distinguish two main categories regarding the residuals we obtained: 15 objects (51 Eri b, $\beta$ Pic b,GQ Lup B, HD1160 B, HD4747 B, HD95086 b, HIP65426 b, HIP107412 B, HR2562 B, HR3549 B, PDS 70 b, PZ Tel B, HR8799 c, d, and e) are located in the innermost part of the system (within $ 1''$), in regions where speckle contaminants are very bright and mostly dominate the residuals once the companion is subtracted. 
We show in Figure \ref{HIP107412} an example of this kind of system, HIP107412 B, with the brown dwarf placed at $0.27"$. The residuals are clearly dominated by bright speckles that cover any fainter feature around the companion. 

\begin{figure} [h!]
\centering
\includegraphics[scale=0.3]{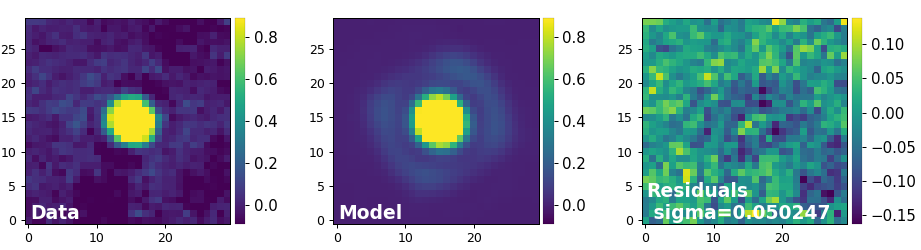}
\caption{cADI image of HR8799 b, the model, and the residuals}
\label{HR8799ADI}
\end{figure}

The other 12 companions are instead orbiting far from the central star and the speckle noise is considerably lower. We make further distinctions between three faint companions (GJ504 B, HD19467 B, and HR8799 b), for which the subtraction proves effective (see Figure \ref{HR8799ADI}), and nine bright companions (DH Tau B, CT Cha B, Eta Tel B, HIP78530 B, HIP64892 B, AB Pic B, TYC 8047- 232-1 B, TYC 7084-794-1 B, and HD 1160 C), for which spurious features may appear in the residuals. Indeed,  it may happen that speckles generated around low contrast companion are bright enough to be above the background. Since we are performing post processing techniques with the cube centered on the star, speckles around off-axis PSFs  are not cancelled and may contaminate the residuals. This kind of issue does not emerge closer to the star because the stellar speckle contaminants dominate there. A second and more frequent bias emerges due to the variation of the PSF during the observation or elongation of the PSF placed close to the edge of the FoV (astigmatism).  Thus, we performed further checks, such as testing different model PSF and different apertures.

\begin{figure} [h!]
\centering
\begin{tabular}{@{}c@{}}
\includegraphics[scale=0.35]{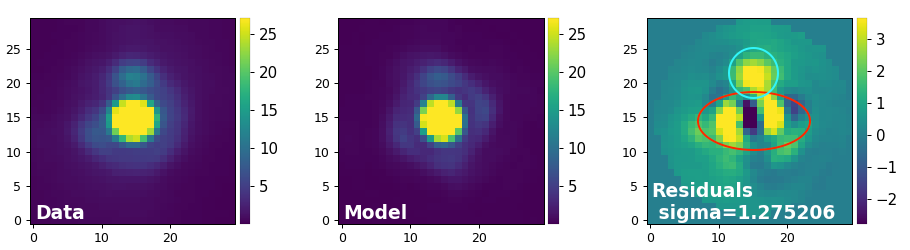}
\end{tabular}
\begin{tabular}{@{}c@{}}
\includegraphics[scale=0.35]{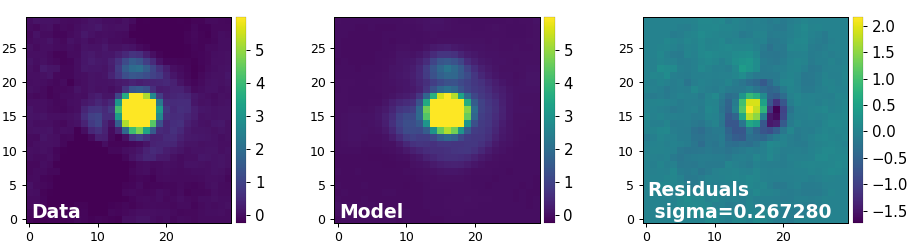}
\end{tabular}
\caption{Upper panel: cADI image of HD1160 C, the model and the residuals; Bottom panel: cADI image of HD1160 B, using HD1160 C as model and the processed image of the subtraction of the two.}
\label{hd1160}
\end{figure}

\begin{figure} [h!]
\centering

\includegraphics[scale=0.43]{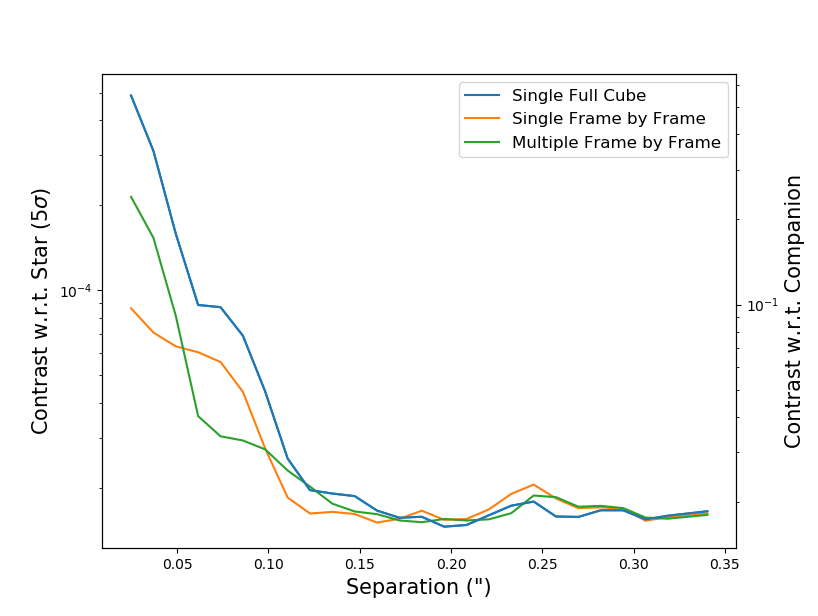}
\caption{Contrast curve around HD1160 B using one single model PSF with the \textit{single\_fullcube} (blue curve) and \textit{single\_framebyframe} (orange curve) routines, and multiple model PSFs using HD1160 C in place of the star for the \textit{multiple\_framebyframe} routine (green curve). Contrasts are calculated with respect to the central star on the left y-axis and with respect to HD 1160 B on the right one.}
\label{hd1160cc}
\end{figure}

As an example of the analysis performed to further investigate the nature of the residuals, we show in Figure \ref{hd1160} the results for HD1160 C and HD1160 B. In the residuals of HD1160 C (last panel of the first row of Figure \ref{hd1160}), a number of structures are present, such as the bright spot in the north of the image (light-blue circle) and the halo close to the center (red ellipse).
In order to obtain the second row of Figure \ref{hd1160}, we used the \textit{multiple\_framebyframe} routine on HD1160 B using as multiple model PSF HD1160 C. The halo  is still present in the residuals, as a negative excess in this case, whereas the bright spot is canceled. The latter is indeed associated to an effect that is, in turn, due to low wind conditions during the observations \citep{milli2018}, while the presence of the disk-like feature is due to an elongation of the PSF of the C companion, which is as expected given its large separation from the host star. In Figure \ref{hd1160cc}, we also show the contrast curves (see Section 2.2) obtained for HD1160 B when applying the \textit{multiple\_framebyframe} routine using the other companion in the FoV with respect to the results obtained with the \textit{single\_framebyframe} routine and \textit{single\_fullcube} with the star used as model PSF. The contrasts achieved here are calculated both with respect to the star (left y-axis) and to HD 1160 B (right y-axis).

A first gain is obtained when considering the \textit{single\_framebyframe} with respect to the \textit{single\_fullcube} due to variable weather conditions during the observation. The further gain in contrast with the \textit{multiple\_framebyframe} routine is evident between $ 0.05''$ and $0.1'',$ where the bright spot in Figure \ref{hd1160} is suppressed, whereas the subtraction is slightly worse closer to the central peak due to the difference in shapes between the two PSFs. \\

\begin{figure} [h!] 
\centering
\begin{tabular}{@{}c@{}}
\includegraphics[scale=0.3]{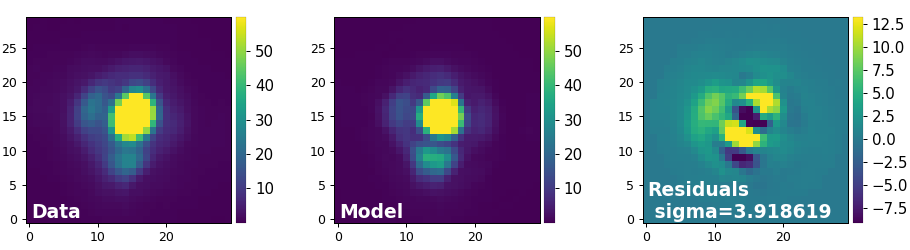}\\
\end{tabular}
\begin{tabular}{@{}c@{}}
\includegraphics[scale=0.3]{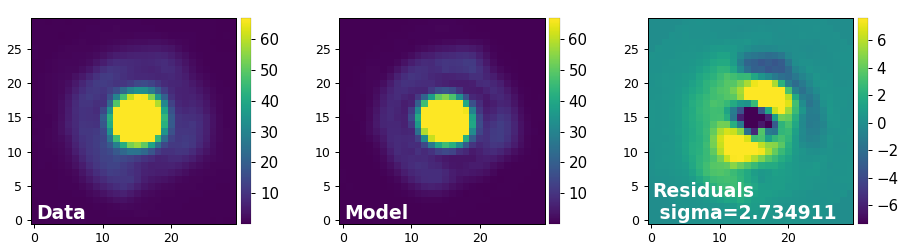}\\
\end{tabular}
\caption{cADI image of TYC 7084-794-1 B, the model and the residuals for the first (upper panel, H band) and the second (bottom panel, K band) epochs.}
\label{TYC 7084}
\end{figure}

\begin{figure} [h!] 
\centering
\begin{tabular}{@{}c@{}}
\includegraphics[scale=0.3]{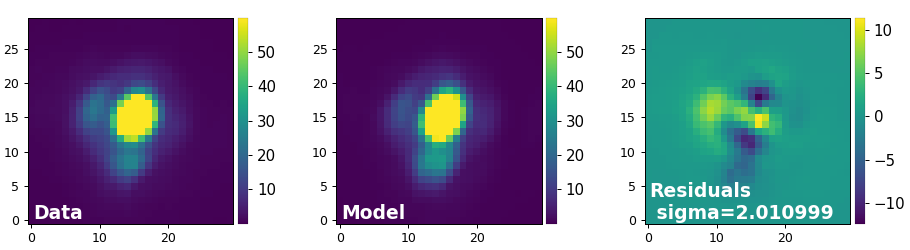}\\
\end{tabular}
\begin{tabular}{@{}c@{}}
\includegraphics[scale=0.3]{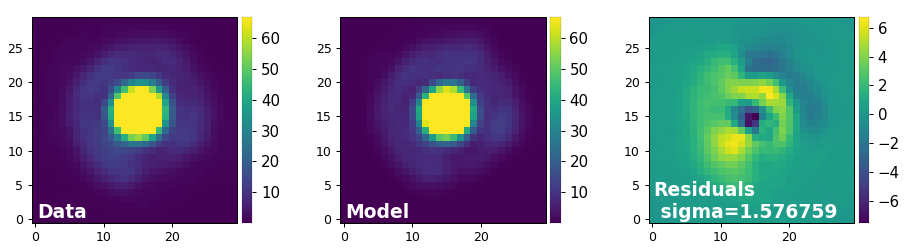}\\
\end{tabular}
\caption{cADI image of TYC 7084-794-1 B, with the model taking into account the smearing of the PSF of the companion and the residuals for the first (upper panel, H band) and the second (bottom panel, K band) epochs.}
\label{smearing}
\end{figure}

We show in Figure \ref{TYC 7084} another example of extended structure obtained around, TYC 7084-794-1 B. These results are particularly misleading because they appear in two different observations, and both in H and K, as a disk-like shape as shown in Figure \ref{TYC 7084}. However, the extended features are produced by a rotational smearing. This happens any time the exposure time for singular frames is too long with respect to the rotation of the objects in the FoV. Usually, this effect is very small and negligible, with the exception of a few systems that have to be treated individually. TYC 7084-794-1 is an example of these since it is very close to the zenith at meridian passage from Paranal and with a detector integration time (DITs) of only 64s, the brown dwarf, placed at a separation of $2.99''$, changes its position by roughly five pixels. For this reason, the PSF of the object is elongated with respect to the PSF of the star. One possible way to overcome this kind of effect might be to elongate the model PSF in the same way as the PSF of the object or to use a reference PSF in the FoV at similar separation, if any. We show in Figure \ref{smearing} the sensitively improved subtraction obtained after considering a PSF model elongated in the same direction of the BD. 

Based on the analysis performed on a few test companions, it emerged that structures in the residuals must be treated very carefully in order to discriminate between those that are artifacts and those that are real. Starting from the end of 2017, one of the greatest source of artifacts in the residuals, the low wind effect, is almost completely fixed, with a number of affected nights at the level of $\sim3$\% \citep{milli2018}. 

Useful information about the observational parameters are contained in the SPARTA files that are produced together with the scientific data by the SPHERE Data Center. These files provide information on how the wind velocity and the seeing varied during the observation for each frame. Moreover, non coronagraphics images of the star collected with the differential tip-tilt sensor (DTTS, see Appendix B) in the H band can be used to test the quality of the PSF during the time sequence (see Appendix B for a detailed description). Thanks to this supplementary information, we can determine, with higher confidence, the nature of the residuals that we obtain.\\

\section{DH Tau B}
Among the targets we analyzed, the most convincing case  is represented by 
a candidate companion close to DH Tau B. 
\subsection{The DH Tau system}
DH Tau is a young (age $\sim$ 1.4 Myr, see Table \ref{tabu2}) low-mass star in the Taurus star-forming region.
DH Tau is still in the accreting phase and forms a wide binary (15" projected separation) with a similar star, DI Tau.
A low-mass brown dwarf or massive planet (DH Tau B) was discovered by  \citet{Itoh}.
Its mass is estimated to be close to the deuterium burning limit.
DH Tau B was shown to be accreting from prominent $H{\alpha}$ and Paschen$\beta$ emission 
\citep{Zhou,Bonnefoy3}.
ALMA observations by \citet{Long} resolved a faint and compact disk around the central star, while no disk around DH Tau B was detected \citep{Wu,Perez}.

\subsection{Observations}
The DH Tau system was observed across multiple epochs with SPHERE, on the dates 2015-10-26, 2015-12-19, and 2018-12-16 within the guaranteed time of the SPHERE Consortium in IRDIFS mode (H2H3 filters) and on the date 2020-01-05 after the acceptance of an open time proposal (0104.C-0327(A)) in IRDIFS\_EXT mode (K1K2 filters). The SPHERE data were reduced as described in Section 3. We also analyzed the SPHERE differential tip-tilt sensor (DTTS) images, which show a reconstruction of the PSF of the star behind the coronagraph (see Appendix B). With this supplementary information, we can evaluate whether the PSF was affected by consistent variations in atmospheric conditions and, thus, if potential structures detected in the residuals are more likely to be real features or spurious artifacts. 

The system was also observed with LBTI using the LMIRCam instrument in L' band ($\lambda=3.6 \mu m $) on the date 2019-09-19. More details about LBTI/LMIRCam and the data reduction performed are given in Appendix C.

\subsection{The candidate companion}
\begin{figure} [h!] 
\centering
\begin{tabular}{@{}c@{}}
\includegraphics[scale=0.3]{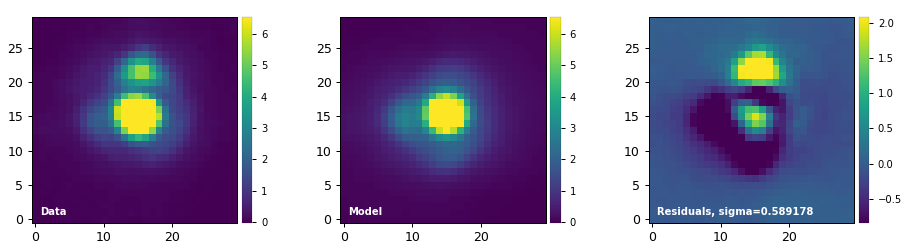}\\
\end{tabular}
\begin{tabular}{@{}c@{}}
\includegraphics[scale=0.3]{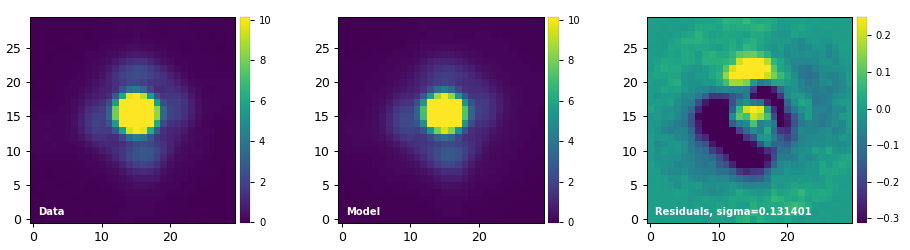}\\
\end{tabular}
\begin{tabular}{@{}c@{}}
\includegraphics[scale=0.24]{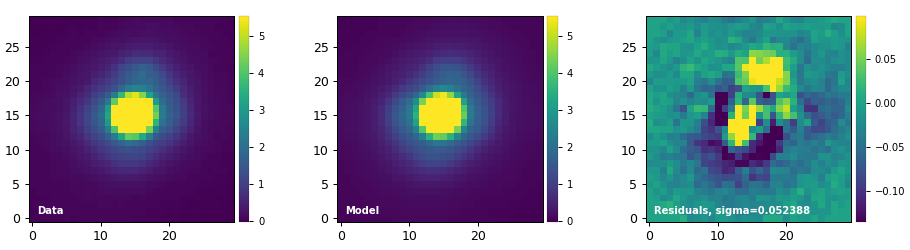}\\
\end{tabular}
\begin{tabular}{@{}c@{}}
\includegraphics[scale=0.3]{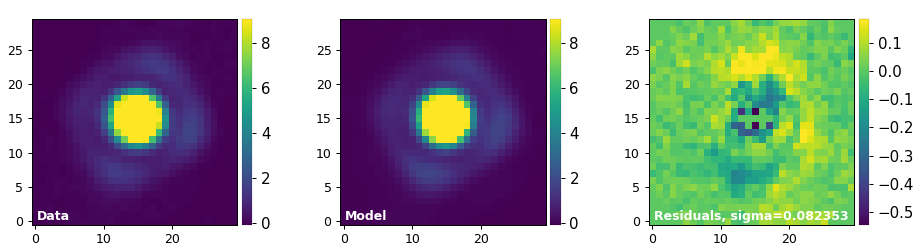}\\
\end{tabular}
\begin{tabular}{@{}c@{}}
\includegraphics[scale=0.3]{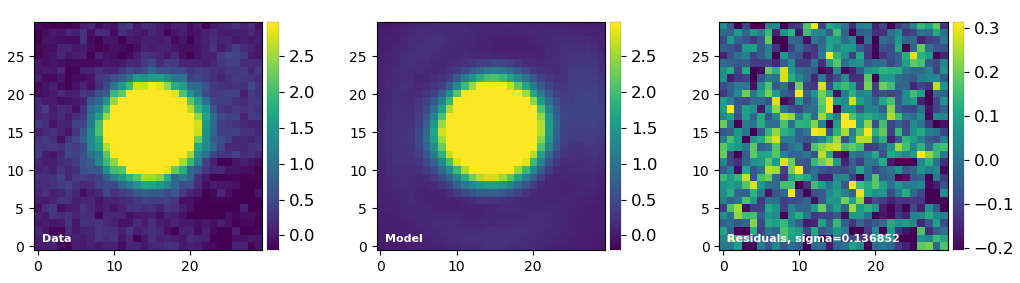}\\
\end{tabular}

\caption{From the top, epoch 1 (2015-10-26), epoch 2 (2015-12-19), epoch 3 (2018-12-16) in H2-band, and epoch 4 (2020-01-05) in K1-band for DH Tau B as observed with SPHERE/VLT, and the observation (2019-09-19) in the L'-band obtained with LMIRCam/LBTI. The candidate companion is visible in the right panels of the SPHERE observations. There is no detection for the LBT data.}
\label{DHTau}
\end{figure}

The detections were initially retrieved from the first two SPHERE datasets listed in Table \ref{tabu1}. The results are shown in the top two rows of Figure \ref{DHTau}. 

To confirm the detection, DH Tau was re-observed with SPHERE with the H2H3 and K1K2 filter pairs. With regard to the H-band observation, the candidate was detected in the residual at a similar separation and position angle with respect to the previous observations (third row of Figure \ref{DHTau}). In the K-band, instead, the candidate is resolved only in K1 (fourth raw of Figure \ref{DHTau}), whereas in K2, a marginal excess is detected in the residuals even if it is not significantly above the background.

To characterize the candidate, which we call DH Tau Bb, we discarded the first epoch due to poor observational conditions and a strong low-wind effect. In Figure \ref{Dtts}, we show, as an example, the DTTS images of the PSF of the central star at different times during the first and second observations taken with SPHERE (see Appendix B). In the first epoch, there are constant variations to the shape of the PSF almost throughout  the entire observation. During the other observations (and, similarly, during the third and fourth datasets), there are no significant differences between the PSFs, thus strengthening the detection of the candidate companion. 

The error bars on the position and contrast of DH Tau Bb were evaluated empirically removing a model of the secondary companion from the raw data, and injecting fake secondary companions at similar contrast and angular separation from the brown dwarf. We then retrieved the contrast and position of the injected fake companions with the tools described in Section 2, and evaluated the mean error on the retrieval. 

As mentioned above, a disk was detected around the central star with ALMA  observations \citep{Long}. Since the disk was very faint in the ALMA observations and was not detected in scattered light using our own SPHERE observations, we do not expect it to impact the PSF subtraction.
The parameters retrieved for DH Tau Bb are listed in Table \ref{tabu3}.

\begin{table*}
\caption{Parameters for the candidate DH Tau Bb}
\label{tabu3}   
\centering
\begin{threeparttable}
\begin{tabular}{c c c c c c c}
\hline\hline
Instrument & Epoch & Band & Contrast & Mass & Separation & Pos Angle\\
      &     &   & &($M_{jup}$)&('')&($^{\circ}$)\\
\hline
\\  
SPHERE &2015-12-19 & H2 & 10.82$\pm$ 0.18  & 0.99$\pm$0.02 & 0.077$\pm$ 0.004 & 4.5$\pm$ 3.5 \\[0.75ex]   
       &2015-12-19 & H3 & 10.78$\pm$ 0.17  &1.03$\pm$0.02 & 0.082 $\pm$ 0.005  & 4.3 $\pm$5.2 \\[0.75ex]
      &2018-12-16 & H2 & 10.85$\pm$ 0.11  &  0.98$\pm$0.01 & 0.078 $\pm$ 0.007  & -10.5 $\pm$ 3.4 \\[0.75ex]  
       &2018-12-16 & H3 & 11.00$\pm$ 0.14 &1.00$\pm$0.01 & 0.086 $\pm$ 0.005  & -7.9 $\pm$ 4.2 \\[0.75ex]
       &2020-01-05 & K1 & 11.37$\pm$ 0.21  & 1.01$\pm$0.03 & 0.100$ \pm$ 0.003 & -4.2$\pm$2.4 \\[0.75ex]
       &2020-01-05 & K2 & $\ge$10.4  & $\le$1.64 &  &  \\[0.75ex]
LBT    &2019-09-19   & L' & >9.2 & <2.6    &   &
\end{tabular}
\end{threeparttable}
\end{table*}
We did notice a drift in the separation of the PSF of the candidate from the H2 to the K1-band. This might indicate that the candidate is not real since it changes its position roughly proportionally to $\lambda/D$. On the other hand, the change in position might be explained by a combination of other factors. First of all, the PSF in the K-band is much broader than in the H-band so that it is more difficult to disentangle the two objects. Moreover, we notice that the position of the candidate is coincident with the position of the first Airy ring of the central companion. Thus, the superimposition of the flux of DH Tau Bb with the flux coming from the  Airy ring of the brown dwarf, which is not well-subtracted, might cause a shift in the peak of the candidate PSF. Additionally, the background at increasing wavelengths is more noisy so that the signal to noise ratio is worse in the K-band than in the H-band. 

The position angle of the candidate is even more difficult to determine with high precision since, at the separation of the candidate, the uncertainty of one pixel corresponds to $\sim 10^{\circ}$. We obtained different values for each epoch and wavelength, with a maximum variation of $\sim 15^{\circ}$ between the second and third epochs. This might be partially due to the orbital motion of DH Tau Bb around the brown dwarf. In fact, assuming a face-on system and a circular, coplanar orbit for the candidate around the brown dwarf, we would obtain a period of $\sim 320 yrs$ for a radial separation of $\sim 0.08''$. In the four-year baseline, the candidate companion should then change its position angle by $\sim 4^{\circ}$, or more if the orbit is not circular and the DH Tau Bb is close to periastron passage. Thus, taking into account the orbital motion, the unknown inclination of the system and the error bars for each position angle, the values obtained are reasonably consistent. Moreover, the time baseline between the epochs allows us to rule out the (a priori\textit{} extremely unlikely) possibility of a background object projected by chance very close to DH Tau B, considering the proper motion of the central star ($\mu_{\alpha}$: 7.07$\pm$0.11 ; $\mu_{\delta}$: -20.70$\pm$0.08 from Gaia DR2). This would imply a shift of about 21 mas toward west and 61 mas toward North of a stationary background source with respect to DH Tau B between epochs 2 and 3. 

\begin{figure*} 
\centering
\begin{tabular}{@{}c@{}c@{}}
\includegraphics[width=9.cm]{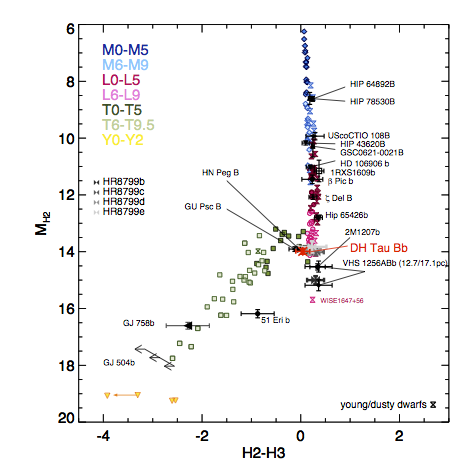}\\
\end{tabular}
\begin{tabular}{@{}c@{}}
\includegraphics[width=9.cm]{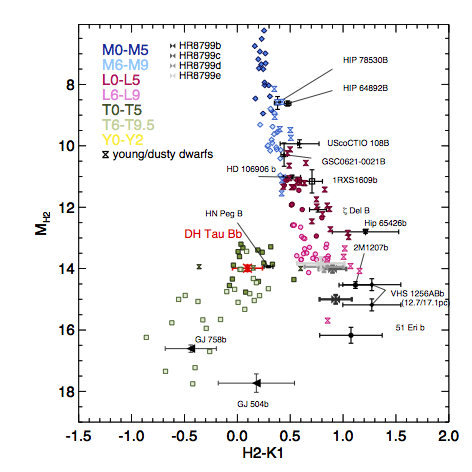}\\
\end{tabular}
  \caption{Color-magnitude diagrams for the IRDIS/SPHERE photometry for the candidate DH Tau Bb compared to a collection of objects from the literature \citep{Bonnefoy2}.}
  \label{CMD}
\end{figure*}

The last parameter retrieved was the contrast. Adopting the age of $1.4$ Myrs, we then calculated the mass of the candidate using the  BT-SETTL models. The values obtained for the mass of DH Tau Bb, $\sim 1$ M\textsubscript{Jup}, are consistent at each epoch and in each filter at which the candidate was detected (H2, H3 and K1). The colors H2-H3 and H2-K1 of the candidate point towards an early T-type, as shown in Figure \ref{CMD}, where we compare them with those for a collection of other low-mass objects.  \citep{Bonnefoy2}.

In order to confirm the detection and better constrain the spectral characteristics of DH Tau Bb, we observed the system in L'-band with the LMIRCam/LBTI instrument (see Appendix C). The candidate was not detected in the residuals, as shown in the last panel of Figure \ref{DHTau}.
Therefore, we determined only an upper limit to the contrast of the object in L', $\sim 9.2$ , which corresponds to an upper limit in mass of $2.6$ M\textsubscript{Jup}. This is consistent with the mass of DH Tau Bb as obtained in H and K band with SPHERE. The non-detection in L'-band strengthens the hypothesis that the companion, if real, is an early T-type object rather than a late L spectral-type that would have been more common considering the young age of the system. Indeed, from the color-magnitude diagram, an L-type companion is expected to be brighter in the L'-band, with a gain in $\Delta mag \sim 1-2$ with respect to the H-band. The corresponding contrast would have been $\sim 3-6 \times 10^{-4}$ that is (marginally) detectable in the observations taken with LMIRCAM/LBTI.  

The DH Tau system has been observed with many different instruments since its discovery \citep{Itoh}. Therefore, we analyzed archive data taken with the WFC3/Hubble Space Telescope (HST) and with the NIRC2/Keck Telescope. For both observations, we did not detect any signs of binarity (the results are omitted from this paper). Indeed, we did not expect to find prominent features either in reflected light or in the poor-quality L' dataset available.

\subsection{Discussion}

The observations described in this paper suggest the presence of a planetary-mass companion ($1$ M\textsubscript{Jup}) at 10 au around DH Tau B.
We consider the object as candidate rather than a confirmed companion because of the small wavelength-dependence of the separation and the instability of the PSF affecting some of the observations.
On the other hand, the roughly consistent position and flux of the candidate in four different datasets spanning several years argues against an instrumental artifact and it is unique among the targets considered in our study. 

This would be the first case of a planetary-mass companion around a brown dwarf bound to a star. If the candidate is confirmed, the DH Tau system will become the first laboratory to test formation mechanisms for binary planet and brown dwarf systems. While a detailed study of the possible formation scenarios is not warranted at this stage, we briefly discuss some possibilities below. 

We may confidently exclude the formation of DH Tau Bb by core accretion in a disk surrounding the brown dwarf due to the large masses involved and the wide separation. In order to generate a secondary gas giant, the circumplanetary disk would have to be unrealistically massive with respect to the mass of DH Tau B.

Some disk instability models showed outcomes similar to those observed here \citep{Stamatellos} but with very low probability.
However, more recent models of disk instability \citep{Forgan} fail to reproduce the architecture observed for the DH Tau system. 
Considering the low mass of the central star ($0.37M_{\odot}$), this mechanism does not appear as a plausible explanation.
Another mechanism that could explain the existence of this system configuration is presented in \cite{Ochiai}, where the authors assess that the formation of planet-planet pairs should be a quite efficient mechanism, with a $10\%$ success rate for planets that undergo orbital crossing. Therefore, it might be possible that DH Tau B and DH Tau Bb formed independently in the circumstellar disk of DH Tau and then go through orbital crossing and close encounters, forming a binary system. However, 
the dynamical simulations by \cite{Ochiai} predict that pairs of planets formed in this way shrink to a very close separation through tidal dissipation. The wide separation observed might be explained in this framework if we manage to catch the high eccentric phase soon after the encounter. 

The latter scenario considers the system to have formed as a low-mass triple system through cloud fragmentation. The mass ratio of $0.027$ between DH Tau B and DH Tau and their wide separation is consistent with this mechanism. On the other hand, companions to brown dwarfs are typically found to have mass ratios
close to unity, both for isolated brown dwarfs in the field \citep{Fontanive1} and for the few cases of bound brown dwarfs (see Section 1). The $0.1$ mass ratio between DH Tau Bb and DH Tau B 
would, then, appear quite unusual. A $1$ M\textsubscript{Jup} object is also expected to be small for cloud fragmentation, being close or below the minimum mass for this process that is quoted in the literature \citep{Boyd,Whitworth,Zuckerman}.

The orbital crossing and cloud fragmentation scenarios, even with the mentioned caveats, seem to be the most likely to explain this peculiar system. Once the candidate is definitively confirmed, dedicated simulations will be needed to investigate these two formation channels in more depth.

An alternative interpretation is that the feature detected in the residuals is a portion of a disk surrounding the brown dwarf. Indeed, accretion through a disc would be consistent with the significant Paschen $\beta$ emission for DH Tau B.
This could explain the shift with the wavelength since our analysis is sensitive to grains of different sizes. Moreover, an extended source would imply large uncertainties in the position angle. However, the disk was not detected in ALMA observations and it was estimated to be very compact around the brown dwarf \citep{Wu,Perez}. This would be inconsistent with our estimate of a separation of $\sim 10$ au.

\section{Statistics of binarity of bound companions}

\begin{figure*} 
\centering
\begin{tabular}{@{}c@{}c@{}}
\includegraphics[width=9.cm]{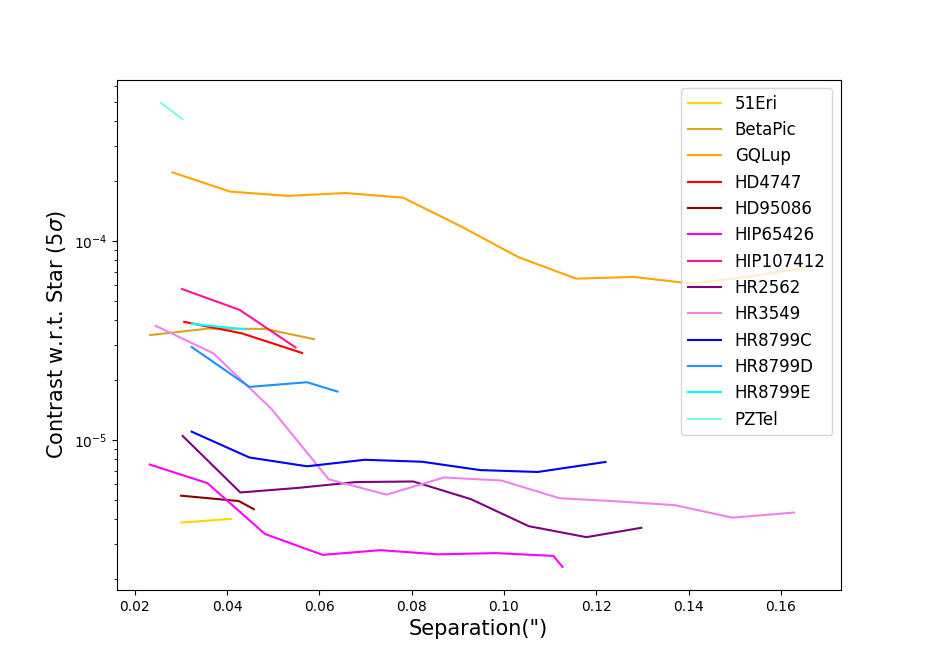}\\
\end{tabular}
\begin{tabular}{@{}c@{}}
\includegraphics[width=9.cm]{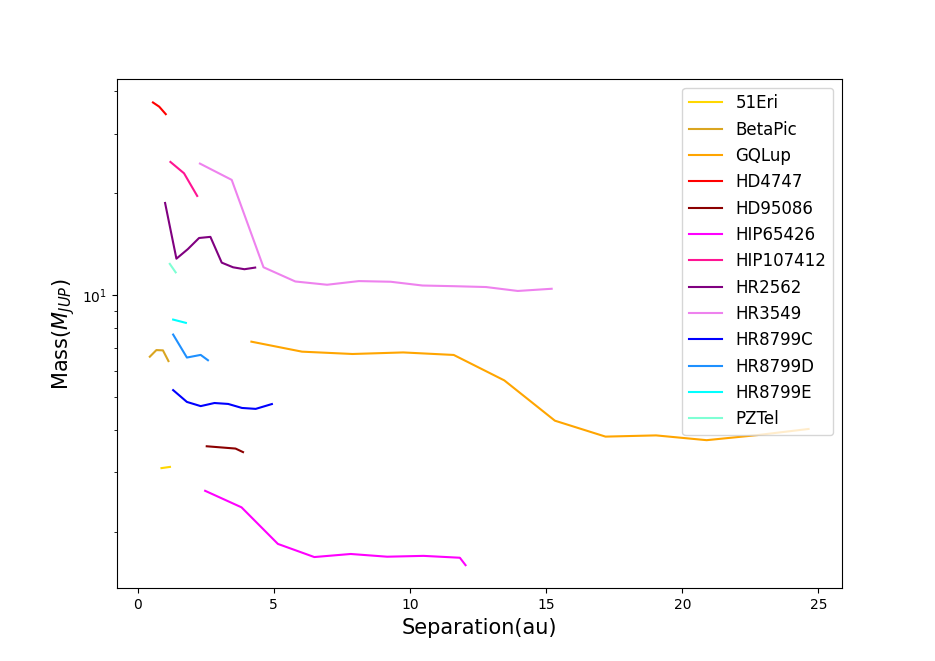}\\
\end{tabular}
\begin{tabular}{@{}c@{}}
 \includegraphics[width=8.9cm]{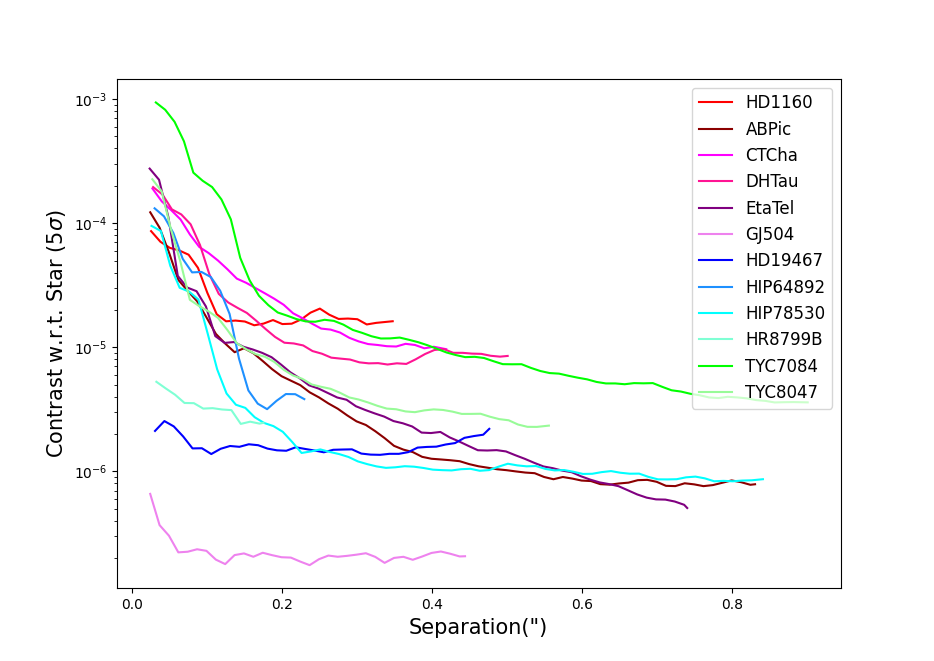}\\
\end{tabular}
\begin{tabular}{@{}c@{}}
 \includegraphics[width=8.9cm]{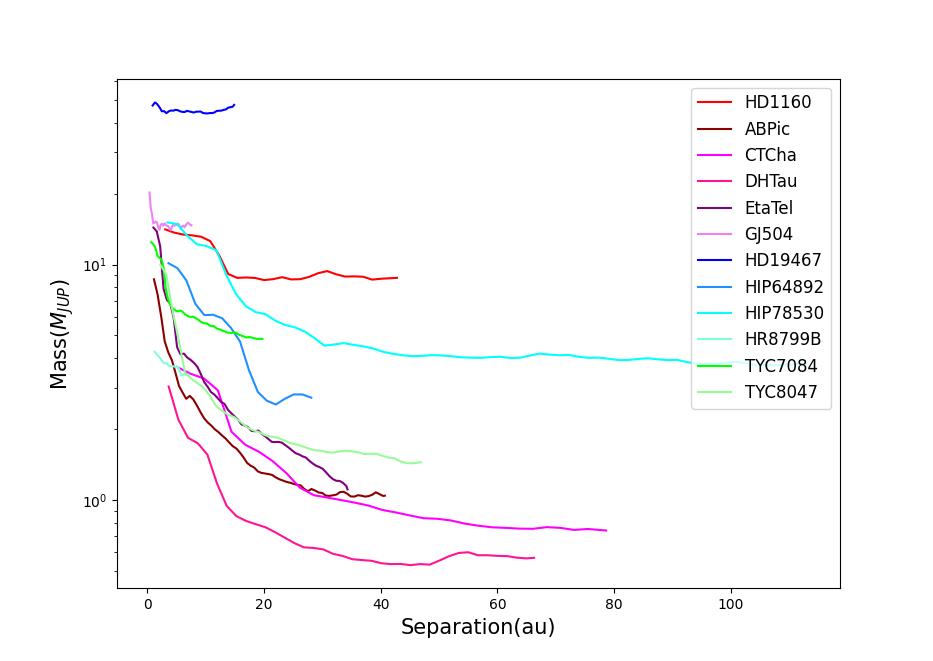}\\
\end{tabular}
\caption{Detection limits for companions placed within 1'' separation from the host star (upper panels) and beyond 1'' (lower panels, HD 1160 B was moved to the second group for a better visualization of the results). Figures on the left show contrasts with respect to the central star as a function of separation (in arcsec); on the right, we show detection limits expressed in mass as a function of separation (in au).}
\label{detlim}
\end{figure*}

\subsection{Detection limits}
For each companion presented in Table \ref{tabu2}, we took the flux that was derived from the observations listed in Table \ref{tabu1} and converted it into mass accordingly with regard to the age adopted for the system (for systems with multiple observations available, we used the ones marked with an asterix in Table \ref{tabu1}). We used the BT SETTL models \citep{Allard} in most cases with the only exceptions of 51 Eri b, HD95086 b, and GJ 504 b for which the ages and contrasts fell out of the boundaries and the AMES COND \citep{Baraffe} model was applied instead.
We then calculated the detection limits around 25 out of 27 companions (we excluded HD1160 C because it falls in the stellar mass range and PDS 70 b because its Hill radius falls inside one FWHM) in order to investigate the contrast reached in their close neighborhoods. 
The detection limits are obtained from a cADI image of the cube where each companion was already subtracted, as described in Section 2.2. 
The contrasts are then converted into masses using the mean ages of the systems listed in Table \ref{tabu2} and the BT-SETTL models for the most of the sample, with the exception of 51 Eri b, AB Pic B, $\eta$ Tel B, GJ 504b, HD 95086 b, HIP 65426 b, TYC 7084-794-1 B, and TYC 8047-232-1 B for which AMES-COND models were used instead since contrasts fell out of the range adopted by BT-SETTL conversions. The Hill radii are then calculated adopting the mean value of the masses in Table \ref{tabu2} and the detection limits are interrupted at those separations.

We show in Figure \ref{detlim} the detection limits for the close-in (upper panels) and wide (bottom panels) companions. Since HD 1160 B is the only close-in companion with a large Hill radius compared to the other planets and brown dwarfs placed within $1"$, we moved it from the first to the second group so that shorter contrast curves are better visualized. From these curves, it emerges how our detection possibilities are limited in the inner part of the system due to the contamination of the star and smaller Hill radii whereas, far from the center, the region of influence around directly imaged companions is background limited and much wider.

\subsection{Statistical results}

We constrain the binary frequency of bound companions using the statistical code described in \citet{Fontanive1}. This Markov Chain Monte Carlo (MCMC) sampling tool was built using the {\tt emcee} \citep{Foreman-Mackey} python algorithm, enabling the derivation of statistical constraints on stellar or substellar populations (occurrence rates, distributions of companions in the parameter space) based on observed surveys (e.g., \citealp{Fontanive1,Fontanive2}).

Binary companions bound to stars have not been studied in depth to date, neither observationally nor theoretically. As a result, the shapes of the companion distributions in mass and separation for such binary systems currently lack any empirical constraints or theoretical predictions. We thus need to rely on assumptions for the underlying population distributions in order to constrain the binary frequency of these bound companions. Given the relatively low masses of our probed planets and brown dwarfs in comparison to isolated, old brown dwarfs in the field, we adopt the mass ratio and separation distributions derived in \citet{Fontanive1} for late-type T5$-$Y0 objects in the Solar neighbourhood. This includes a lognormal distribution in separation with a peak at 2.9 au and a logarithmic width of 0.21, and a power law distribution with index 6.1 in mass ratio (see \citealp{Fontanive1} for details).

\begin{figure} [h!]
\centering
\includegraphics[scale=0.65]{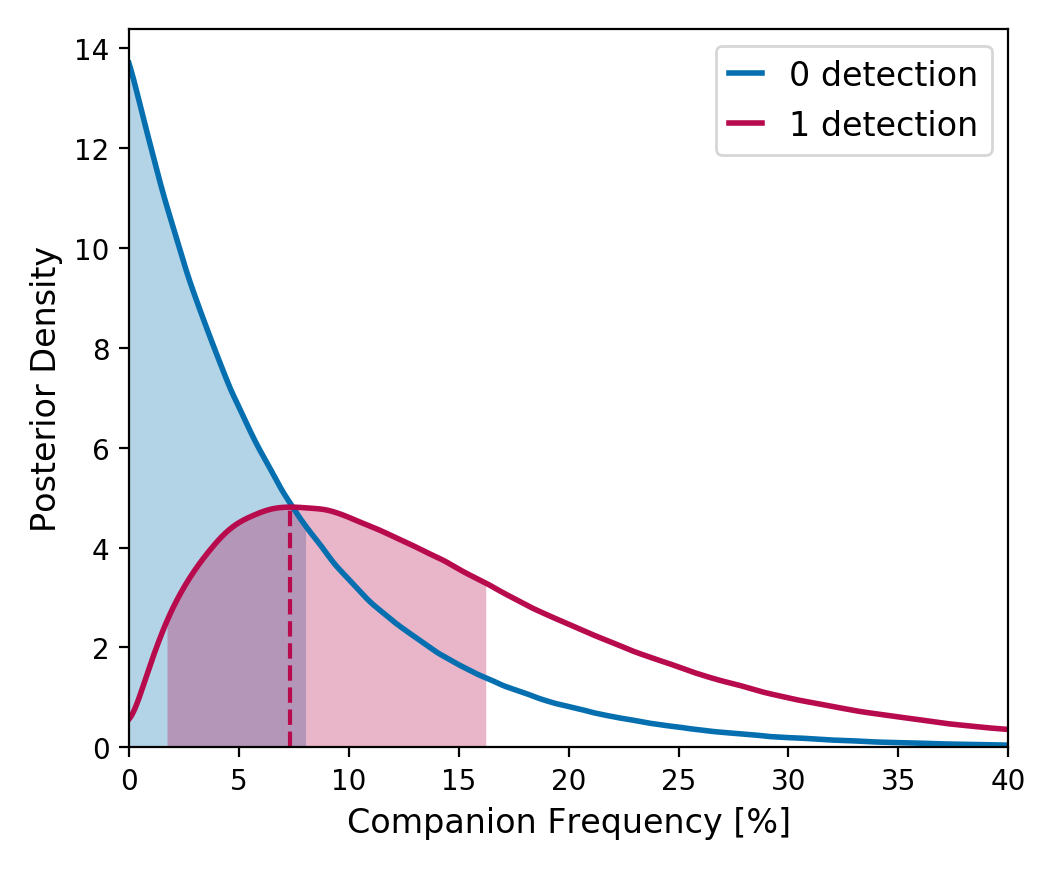}
\caption{Posterior distributions obtained for the frequency of bound binary companions from the MCMC analysis, in cases of 0 (blue) and 1 (red) detection in our survey. Frequencies were derived over separation ranges of 0.5$-$100 au and mass ratios of $q=0.05-1$. The red dashed line shows the peak of the posterior density in the 1-detection scenario, at 7.3\%, while only an upper limit was obtained in the 0-detection case. The shaded areas correspond to the 68\% confidence intervals.}
\label{stat}
\end{figure}

From the calculated detection limits around each target and assuming that those same distributions hold for our surveyed systems, we constrain the binary rate of bound substellar companions between separations of 0.5$-$100 au and mass ratios of $q=0.05-1$. Figure~\ref{stat} shows the obtained posterior distributions from the MCMC analyses performed for respective cases of 0 and 1 detection. As expected, the scenario with no detected companion only provides an upper limit on the binary frequency, of $f<8.1$\% ($<21.2$\%) at the 1-$\sigma$ (2-$\sigma$) level. In the case where our reported candidate DH~Tau~Bb is real, we derive a binary fraction of $f=7.3^{+8.9}_{-5.6}$\% on those mass ratio and separation ranges, which is in good agreement with the results for field brown dwarfs from \citet{Fontanive1} on a comparable region of the parameter space.

We emphasize that the obtained frequencies assume a fixed distribution of companions in the separation-mass ratio space and that these results only hold under those assumptions. While it would be very interesting to explore other possible underlying shapes of companion distributions that may differ from the field brown dwarfs, the existing data do not currently allow for such an analysis. It is worth noting, however, that the one detection present in our observed sample (if confirmed) has a fairly wide separation ($\sim$10 au) and a very low mass ratio ($q<0.1$) compared to the peaks of the adopted field distributions ($\sim$3 au, $q\sim1$). Late-type field binaries indeed rarely have separations larger than around 10 au, and typically show strong preferences for near-equal-mass systems. The disparity between the model and observed populations is further highlighted by the fact that low $q$ values like that of the DH~Tau~B system are only reached around few targets in the SPHERE sample. This suggests that the true binary companion distributions for bound planets and brown dwarfs are likely to be somewhat different than for the field population, although deeper observations of larger sample sizes will be required to confirm this and to further explore formation theories for bound and isolated systems.

\section{Conclusions}
In this paper, we present tools based on the NEGFC technique to suppress the flux of known faint planets and brown dwarfs in order to detect potential companions or disks around them. The different routines are optimized in order to determine, with the best possible precision, the position and the flux of the companion, to subtract it, and to look into the residuals for additional faint features bound to it. We distinguished among two kinds of routine: the \textit{single\_fullcube} and the \textit{single\_framebyframe} routines, which use a single model PSF, and the \textit{multiple\_framebyframe} routine, which uses more PSFs as models to subtract the companion. The single-model PSF tools are usually used for coronagraphic observations, whereas if a suitable PSF is visible in each raw frame the multiple-PSFs routine can be applied (both in coronagraphic and non-coronagraphic datasets). The latter tool is very helpful to suppress spurious features that may appear in the residuals due to the variation of the PSF during the observation. However, PSFs that are bright enough and at a good-enough separation to be used as models are rarely found in the FoV of substellar companions. In non-coronagraphic observations, instead, the central star can be used as a model for each frame but such settings are not common for planets hunting due to the poor contrast that can be achieved.

These tools were first tested on synthetic point sources and disks and on selected on-sky test cubes. We then applied it to the substellar companions imaged by SPHERE within the SHINE survey. This is the first methodical study that aims to look for circumplanetary disks and satellites or, more specifically, for the kind of instrument that we have, that is, pairs of bound planets. Among the objects in the sample, we identified a candidate companion around DH Tau B, with a projected separation of $\sim 10$ au from the brown dwarf and with mass a of $\sim 1$ M\textsubscript{Jup}. The DH Tau system was observed with SPHERE (H2-H3 and K1-K2 bands) in four different epochs and with LMIRCam/LBTI (L' band). The candidate was detected in each of the H-band datasets and in the K1-band. The K2 and L'-band observations did not lead to a detection but they were used to obtain only the upper limits on the mass of the candidate, which is consistent with the previous results. The color-magnitude diagrams point towards an early T-type object.

We derived the contrast curves for each of the 25 suitable substellar companions observed with SPHERE and we compared the statistics of the multiplicity of our bound companions to the one published in \cite{Fontanive1} for isolated objects. If the companion around DH Tau B is real, the frequency of binary substellar companion is $\sim 7.3 \%$, which is consistent with the result obtained for field brown dwarfs.

The techniques presented in this paper will serve as a useful tool for the hunt of companions and extended sources around directly imaged companions in the future. Indeed, in the long-term, we plan to adapt the tools described here to prepare for the next-generation telescopes, such as JWST and ELT.

MIRI$@$JWST will provide high-contrast imaging and IFS imaging from 5.6 to 28.8 $\mu$m. The greatest advantage of this telescope is the stability of the PSF over time. For MIRI, the expected contrast goes from $10^{-4}$ to $10^{-6}$ at separations $>2$ arcsec \citep{Danielski}. Even if the contrast were lower than what is achievable with SPHERE, it will be more efficient for colder and smaller objects due to the longer wavelength coverage.

A dedicated high-contrast imager on an ELT \citep[e.g.,][]{Kasper} is expected to reach contrasts of $10^{-8}$ at 30 mas to $10^{-9}$ at larger separations. These contrasts will allow for the detection of Earth-like planets. With this instrument, which will also provide a very stable PSF over time, exomoons and exorings will be much more efficiently detected. These detections will increase the probability of finding habitable zones even at far distances from the host star, much the same way as the moons of our Solar System are best candidates for the research of life forms.

\begin{acknowledgements}
SPHERE  is  an  instrument  designed and built by a consortium consisting of IPAG (Grenoble, France), MPIA (Heidelberg,  Germany), LAM  (Marseille,  France),  LESIA  (Paris,  France), Laboratoire Lagrange (Nice, France), INAF–Osservatorio di Padova (Italy), Observatoire  de  Geneve  (Switzerland),  ETH  Zurich  (Switzerland), NOVA  (Netherlands),  ONERA  (France)  and  ASTRON (Netherlands) in  collaboration  with ESO. SPHERE was funded by ESO, with additional contributions from CNRS (France),  MPIA  (Germany),  INAF  (Italy),  FINES  (Switzerland)  and  NOVA (Netherlands).  SPHERE  also  received  funding  from  the  European  
Commission  Sixth  and  Seventh  Framework  Programmes  as  part  of  the Optical  Infrared  Coordination  Network  for  Astronomy  (OPTICON)  under grant  number RII3-Ct-2004-001566 for FP6 (2004–2008), grant number 226604 for FP7 (2009–2012)  and  grant  number  312430  for  FP7  (2013–2016).  
This  work  has  made  use  of  the SPHERE  Data Centre, jointly operated by OSUG/IPAG (Grenoble), PYTHEAS/LAM/CESAM (Marseille), OCA/Lagrange (Nice) and Observatoire de Paris/LESIA (Paris). We thank P.  Delorme  and  E.  Lagadec  (SPHERE  Data  Centre)  for  their  efficient help  during  the  data  reduction  process. 
This work is based on observations collected at LBT Observatory.
The LBT is an international collaboration among institutions in the United States, Italy and Germany. LBT Corporation partners are: The University of Arizona on behalf of the Arizona university system; Istituto Nazionale di Astrofisica, Italy; LBT Beteiligungsgesellschaft, Germany, representing the Max-Planck Society, the Astrophysical Institute Potsdam, and Heidelberg University; The Ohio State University, and The Research Corporation, on behalf of The University of Notre Dame, University of Minnesota and University of Virginia.
We  acknowledge  financial  support from the “Progetti Premiali” funding scheme of the Italian Ministry of Education, University, and Research. A.Z. acknowledges support from the FONDECYT Iniciaci\'on en investigaci\'on project number 11190837. We  acknowledge  financial  support  from  the  Programme National de Planetologie (PNP) and the Programme National de Physique Stellaire (PNPS) of CNRS-INSU. This work has also been supported by a grant from the French Labex OSUG\@2020 (Investissements d’avenir – ANR10 LABX56). The project is supported by CNRS, by the Agence Nationale de la Recherche.
This work has been supported by the project PRIN-INAF 2016 The Cradle of Life - GENESIS-SKA (General Conditions in Early Planetary Systems for the rise of life with SKA).
We also acknowledge the use of the Large Binocular Telescope Interferometer (LBTI) and the support from the LBTI team.
T.H. acknowledges support from the European Research Council under the Horizon 2020 Framework Program via the ERC Advanced Grant Origins 83 24 28. 
We also want to thank the referee for the useful comments.
\end{acknowledgements}

\bibliographystyle{aa}
\bibliography{bibliography}

\begin{appendix}

\section{Data analysis}

 \subsection{Synthetic companions}
In order to understand how much we can gain from the subtraction of the companion in term of detection limits and which kind of point sources we may expect to detect, we simulated a companion (C1) in the FoV of TYC 8047-232-1, observed with SPHERE in date 2015-09-25 (see Table \ref{tabu1}). This C1 was injected at a separation of $r=3.9''$ and with contrast of $10^{-3}$ with respect to the central star.

To obtain more realistic residuals, we used two different PSFs, one for the injection of the fake companion and one as a model. At this stage, the differences between the \textit{single\_framebyframe} and \textit{single\_fullcube} routine are not appreciable since the injected PSF does not vary. The detection limits around C1 are calculated on a cADI image obtained before and after the subtraction of the companion using \textit{single\_framebyframe} routines and following the procedure described in Section 2.

\begin{figure} [h!]
\centering
\begin{tabular}{@{}c@{}c@{}}
\includegraphics[width=0.5\textwidth]{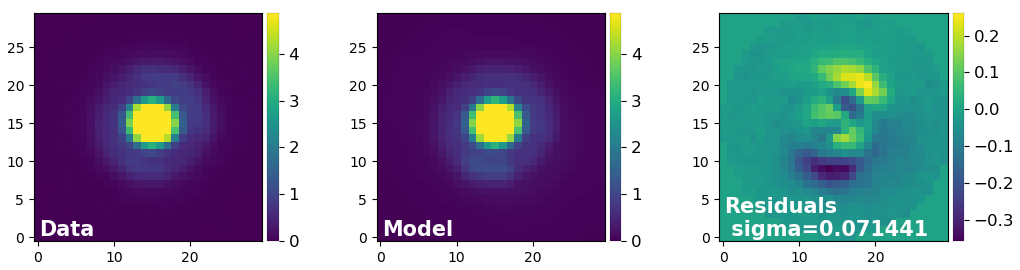}\\
\end{tabular}
\begin{tabular}{@{}c@{}c@{}}
\includegraphics[width=0.5\textwidth]{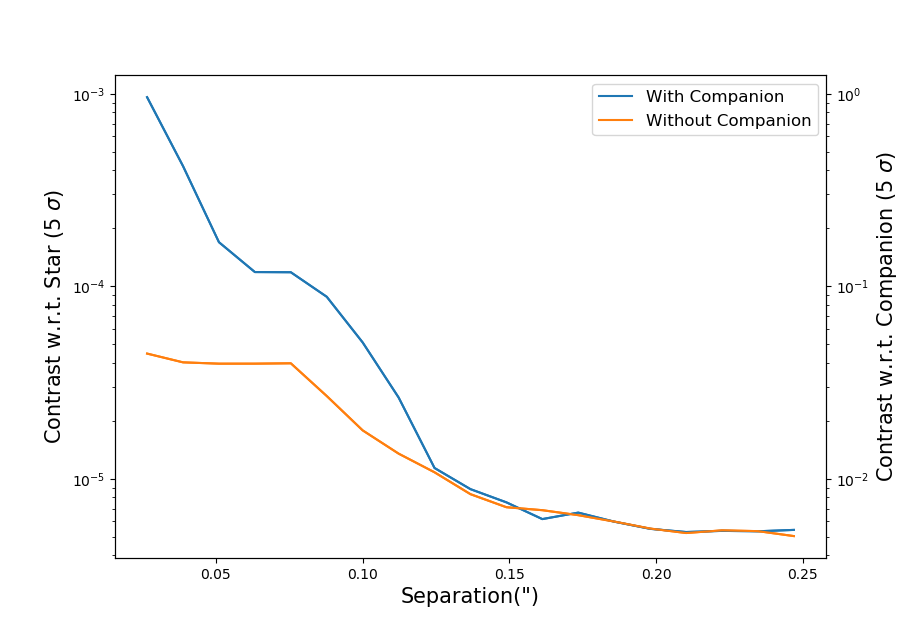}\\
\end{tabular}
\caption{Upper panel: cADI images of, from the left, the companion, the model PSF, and residuals. Bottom panel: Contrast curves centered on the position of the companion as obtained in the post-processed frame with (blue curve) and without (orange curve) the companion. Contrasts are calculated with respect to the star (left y-axis) and with respect to the companion (right y-axis).}
\label{onefake}
\end{figure}

Residuals and contrast curves centered on C1 are shown in Figure \ref{onefake}. The gain in contrast when we get very close to the PSF of the companion appears once the latter is subtracted. This region is also the most interesting to investigate when looking for bound features since, in most cases, it is inside the Hill radii of substellar companions.

\begin{figure} [h!] 
\centering
\begin{tabular}{@{}c@{}}
\includegraphics[scale=0.35]{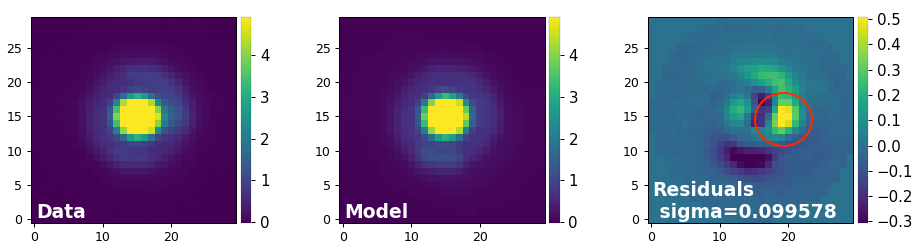}\\
\end{tabular}
\begin{tabular}{@{}c@{}}
\includegraphics[scale=0.35]{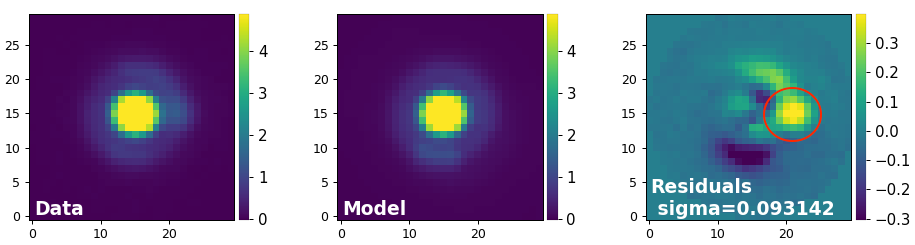}\\
\end{tabular}
\begin{tabular}{@{}c@{}}
\includegraphics[scale=0.35]{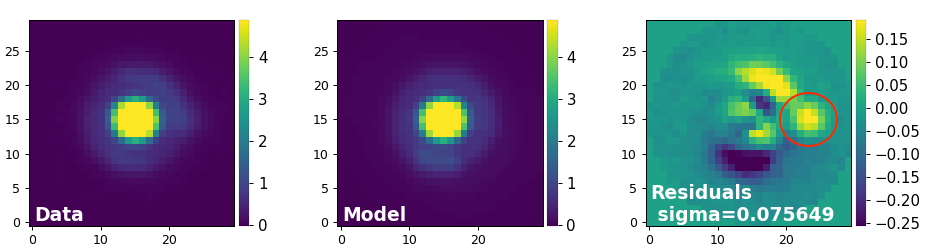}\\
\end{tabular}
\caption{Injection of a second fake companion at the detection limits placed at $0.05''$ (upper panel), $0.075''$ (mid panel), and $0.1''$ (bottom panel) from the peak of the primary.}
\label{twofake}
\end{figure}

In order to simulate two bound substellar objects, we injected very close to C1 a second fake companion (C2) with increasing separations and with contrasts corresponding to the values at detection limits as shown on the contrast curves in Figure \ref{onefake}. The residuals for three different positions are shown in Figure \ref{twofake}. In the upper row, the second companion is placed at $\sim 0.05''$ from the main peak with a contrast of $5\times10^{-5}$, in the mid row at $\sim 0.075''$ and with contrast of $4\times10^{-5}$ and in the bottom panel at $\sim 0.1''$ with contrast of $2\times10^{-5}$. From these preliminary tests, it emerges that our procedure is, in principle, quite efficient in unveiling additional companions placed very close to the PSF of directly imaged planets and brown dwarfs. However, artifacts that are present in the residuals, especially at few $\lambda/D$, may hide or simulate the presence of faint objects. In the upper raw, for example, the fake companion is detected in the residuals even if it could be interpreted as an additional artifact attributed to the PSF subtraction. For such a scenario, more datasets would be needed. Once, instead we move farther from the central peak, the PSF of the secondary is more evidently disentangled by the surrounding residuals.

\subsection{Synthetic disks}
Since the method described in this paper is optimized to subtract the best point-like model to a given PSF, any conspicuous deviation from such a profile should be visible in the residuals whenever it is above the background noise. For this reason, we should be able not only to detect point-like sources but also extended ones. In order to understand which kind of residuals we should expect from extended sources, we inject a circular annulus around the mock planet C1 (see previous Section) with different inclinations.  

Therefore, each raw frame was initially de-rotated into its real position, a one pixel-width annulus with uniform contrast was inclined and injected around the companion. For the forward-modelling process, the planet+annulus were then convolved with the PSF of the instrument. Finally, the frame was rotated again in the original configuration. We must stress that such models are only qualitative and not representative of the physical processes involved (see discussion below). However, more accurate simulations of disks around substellar companions are beyond the purposes of this paper.

\begin{figure} [t] 
\centering
\begin{tabular}{@{}c@{}c@{}}
\includegraphics[width=0.5\textwidth]{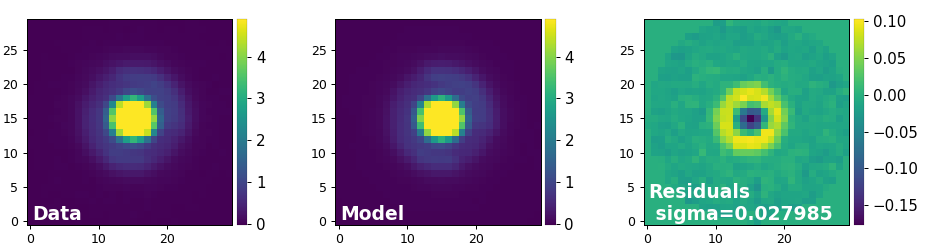}\\
\end{tabular}
\begin{tabular}{@{}c@{}}
\includegraphics[width=0.5\textwidth]{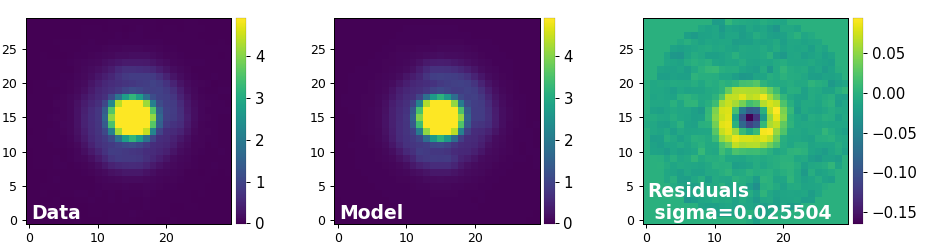}\\
\end{tabular}
\begin{tabular}{@{}c@{}}
\includegraphics[width=0.5\textwidth]{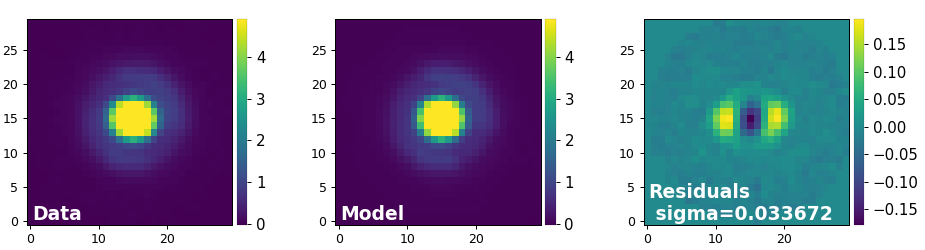}\\
\end{tabular}
\caption{Example of a companion with a synthetic disk with inclination, from top to bottom, of $0^{\circ}$, $45^{\circ}$, $85^{\circ}$}
\label{sindisk}
\end{figure}

We then applied the \textit{single\_framebyframe} routine to the cube with C1 and the synthetic disk. We show in Figure \ref{sindisk} the results obtained. The circular annulus was injected at a separation of $\sim 35 mas$, with an integrated flux ratio with respect to the companion of $F_{disk}/F_{pl}=1/10$. Moreover, we simulated an inclination of the disk moving from a face-on system (upper row), through $45^{\circ}$ of inclination (mid row), to an nearly edge-on configuration ($85^{\circ}$, bottom row). From  Figure \ref{sindisk} it emerges how much the inclination influences the detectability of extended features even in an ideal case like this.

Other factors should be taken into account with regard to the detectability of extended features. Apart from the inclinations that, as shown before, form a critical aspect, also the brightness of the disk and its extension are crucial parameters. In particular, disks around sub-stellar companions are less massive than circumstellar ones. However, their brightness may not be proportional to the quantity of dust grains since they are illuminated both from the inside, by the companion, and from the outside, by the star. Even if disks around planets and brown dwarfs may have higher contrast than expected, we still may not be able to detect them due to the small Hill radii involved. Such compact disks are not resolved and may only be detected by their emission lines.

\subsection{HIP 87836 data}
\begin{figure} [h!]
\centering
\includegraphics[scale=0.5]{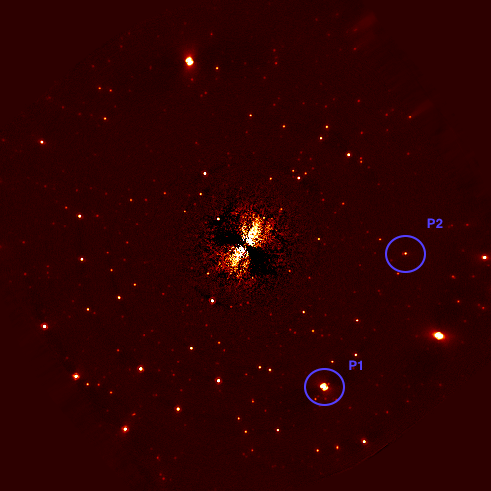}
\caption{cADI image of HIP 87836. Encircled in blue, the two stars used for the analysis are described in Appendix A.3}
\label{adi}
\end{figure}

We also tested the routines described in Section 2 on an ad hoc dataset, HIP 87836. This system is particularly suitable for our purposes since it is placed almost on the galactic plane (Galactic coordinates 001.9927977373370 -01.6464052166128); thus the FoV of the instrument is very crowded with PSFs, as shown in Figure \ref{adi}. Due to the projection on the sky plane, a few of these stars are detected very close to each other so that they may resemble multiple bound systems, similar to the ones that we are investigating. 

The dataset we analyzed comes from observations taken with SPHERE/VLT in May 2018 with the coronagraph on (see Section 3 for the description of the observations and their pre-processing). To test the efficiency of our routines in detecting further companions to directly imaged sources, we chose two apparent multiple systems with different flux ratios and distances between the primary and the secondary that are highlighted by blue circles in Figure \ref{adi}. 

\begin{figure} [t] 
\centering
\begin{tabular}{@{}c@{}c@{}}
\includegraphics[width=0.5\textwidth]{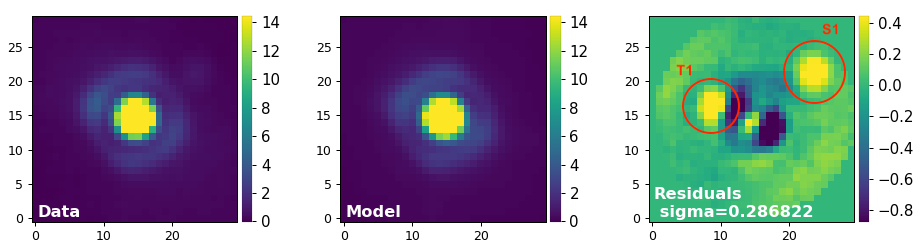}\\
\end{tabular}
\begin{tabular}{@{}c@{}}
\includegraphics[width=0.5\textwidth]{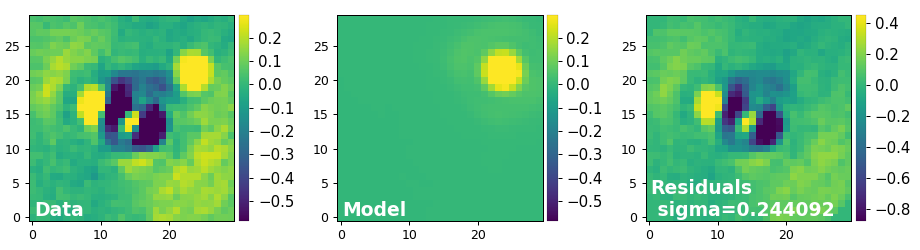}\\
\end{tabular}
\begin{tabular}{@{}c@{}}
\includegraphics[width=0.5\textwidth]{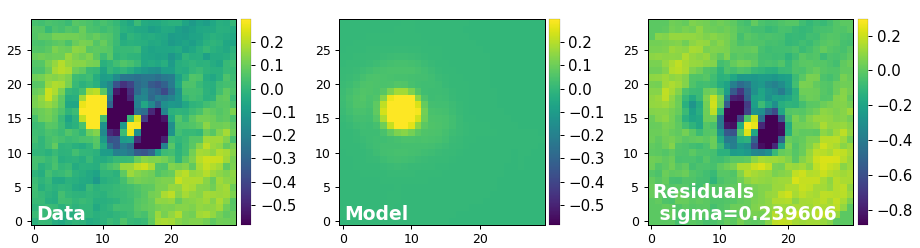}\\
\end{tabular}
\caption{Images of the object P1 around HIP 87836, model and residuals of, from the top, the primary P1, the secondary S1, and the tertiary T1.}
\label{P1}
\end{figure}

The first point source (P1) is placed at $r_{P1}=4.11''$ and $\theta_{P1}=209.1^{\circ}$ and it is very bright, exhibiting a contrast with respect to the central star $F_1=2.5\times10^{-4}$. A second fainter object is placed nearby P1. Our aim is to visualize it in the residuals and, subsequently, to determine its precise position and flux. In order to do that, we ran the \textit{single\_framebyframe} routine to optimize the position and contrast of P1 and, then, to subtract it to obtain the final image shown in the upper panel of Figure \ref{P1}. As expected, the secondary is clearly visible in the upper right corner of the residuals. Moreover, another companion, very close to the primary and not visible by eye in the original image, was unveiled after the subtraction. In order to determine contrasts and positions for the secondary (S1) and the tertiary (T1) we applied twice our routines to subtract each object, in turn. Since S1 and T1 are too faint to optimize $(r, \theta, F)$ in each of the unprocessed frames of the scientific cube, we used the \textit{single\_fullcube} routine. Results of the two subtractions are shown in the second and third rows of Figure \ref{P1}, respectively. The contrast with respect to the primary are very similar, $F_2/F_1=1.7\times10^{-2}$ and $F_3/F_1=1.8\times10^{-2}$, and the spots are placed at $r_{S1}=0.13''$, $\theta_{S1}=305.8^{\circ}$ , and $r_{T1}=0.09''$, $\theta_{T1}=72.8^{\circ}$ with respect to P1. 

\begin{figure} [t] 
\centering
\begin{tabular}{@{}c@{}}
\includegraphics[width=0.5\textwidth]{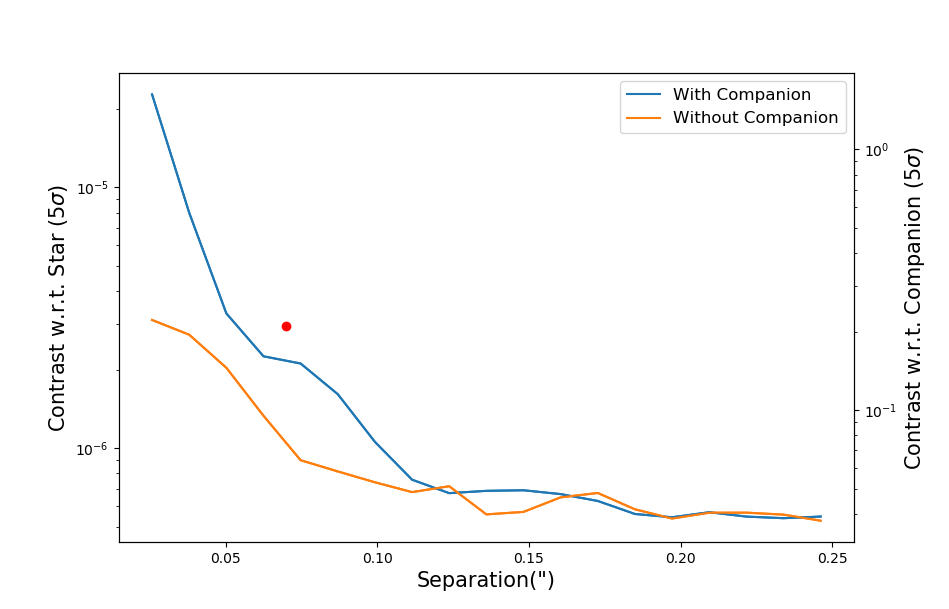}\\
\end{tabular}
\begin{tabular}{@{}c@{}}
\includegraphics[width=0.5\textwidth]{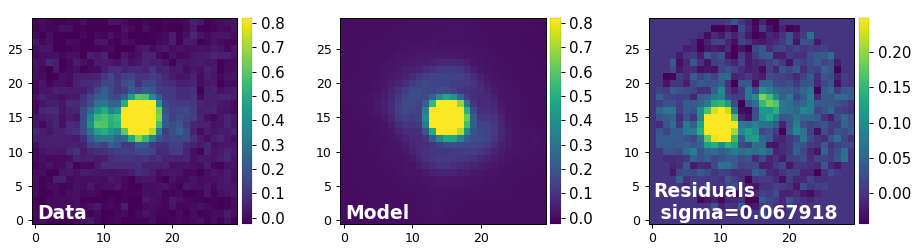}\\
\end{tabular}
\begin{tabular}{@{}c@{}}
\includegraphics[width=0.5\textwidth]{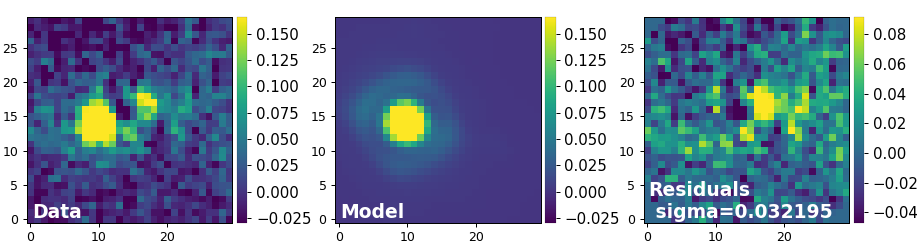}\\
\end{tabular}
\caption{Contrast curve (upper figure) with and without the P2 point source (S2 was removed before calculating the contrasts and is represented by the red dot) and images of the object P2 (lower figures) around HIP 87836, model and residuals of, from the top, the primary P2, and the secondary S2.}
\label{P2}
\end{figure}

The second point source (P2) is placed at $r_{P2}=4.07''$ and $\theta_{P2}=267.1^{\circ}$ and is much fainter than P1, with a contrast of $F_1=1.4\times10^{-5}$. The secondary (S2) is not clearly distinguishable from the primary. Thus, this system is a perfect test bench for our routines with regard to fainter objects with close companions. We ran the \textit{single\_framebyframe} routine to optimize $(r, \theta, F)$ for P2 and to subsequently subtract it, as shown in the upper row of Figure \ref{P2}. In the image of the residuals, S2 is clearly visible and, similarly to what we have done for S1 and T1, we ran the \textit{single\_fullcube} routine to determine position and flux for this object. Results of the second subtraction are shown in the bottom panel of Figure \ref{P2}. The contrast obtained with respect to the primary is $F_2/F_1=2.1\times10^{-1}$ and the spot is placed at  $r_{S2}=0.07''$, $\theta_{S2}=99.3^{\circ}$ with respect to P2. Together with the images of the two point sources, we show in Figure \ref{P2} the contrast curves centered on the position of P2 before (blue curve) and after (orange curve) subtracting P2. Such values were obtained subtracting S2 from each raw frame and, then, using the same procedure described in Section 2. The red dot in the plot represents the position and contrast obtained for S2. 

Unfortunately, we were not able to find any background star that could be used as a model for each frame. Indeed, the only two bright point sources are placed at the very edge of the FoV \textbf{so that} anisoplanatism  affects such PSFs. We tested both objects as model PSFs for the \textit{multiple\_framebyframe} routine but the residuals are higher than the ones obtained with the PSF of the star.

\section{DTTS data}

The differential tip-tilt sensor of SPHERE is a second-stage tip-tilt sensor located just before the near-infrared coronagraphic wheel \citep{Beuzit}. At this location, a grey beam splitter separates a few percent of the light, which is sent to a technical camera sensing precisely the position of the focal spot at a rate of 1\,Hz. This enables the compensation of any slow movement between the coronagraph focus and the visible wave-front sensor due to thermal movements, residual differential dispersion, etc. The differential tip-tilt loop of the AO system ensures the final and accurate centering of the star on the coronagraph. Its precision has been assessed in the integration phase and on-sky to be $<0.5$\,mas, \citep{Baudoz} which means 1/80th of the diffraction width in H-band, for bright stars. For stars as faint in the H-band as DH Tau, the accuracy is probably reduced.

During the observations, the DTTS images are regularly recorded as part of the AO system telemetry. The system records a 32$\times$32-pixel, 1-sec average image every 30 seconds. This series of images enables monitoring different information: flux variation, residual jitter of the PSF, influence of the low-wind effect on the PSF \citep{Sauvage2015LWEAO4ELT,milli2018}. It is, however, important to note that for faint stars, only the tip of the PSF will be visible in the DTTS images due to the faintness of the star combined with the small fraction of the flux that is taken for the tip-tilt sensing. Therefore, mild low-wind effect may not necessarily be detected using the DTTS images.

\begin{figure*} [h!] 
\centering
\begin{tabular}{@{}c@{}}
\includegraphics[scale=0.12]{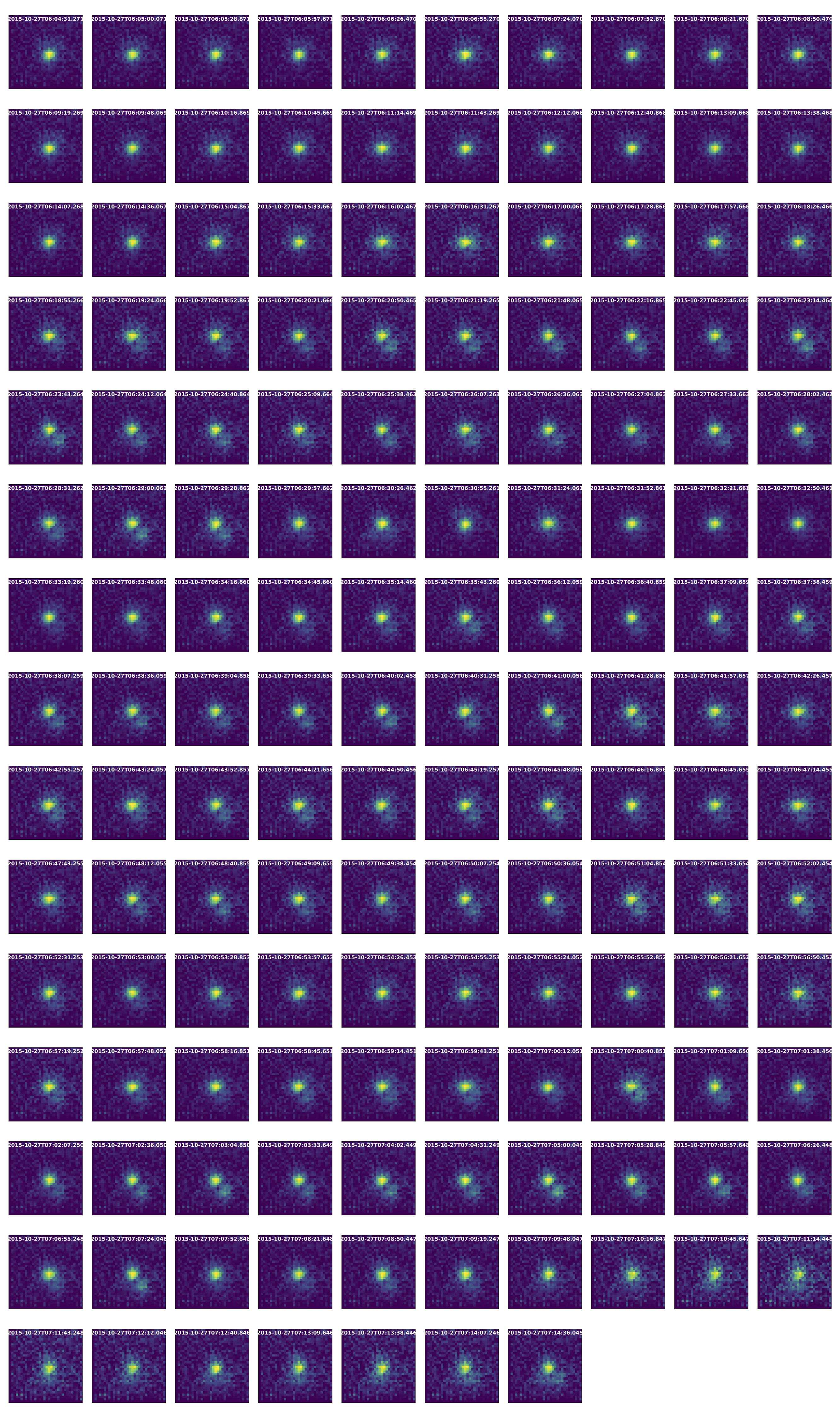}
\end{tabular}
\begin{tabular}{@{}c@{}}
\includegraphics[scale=0.11]{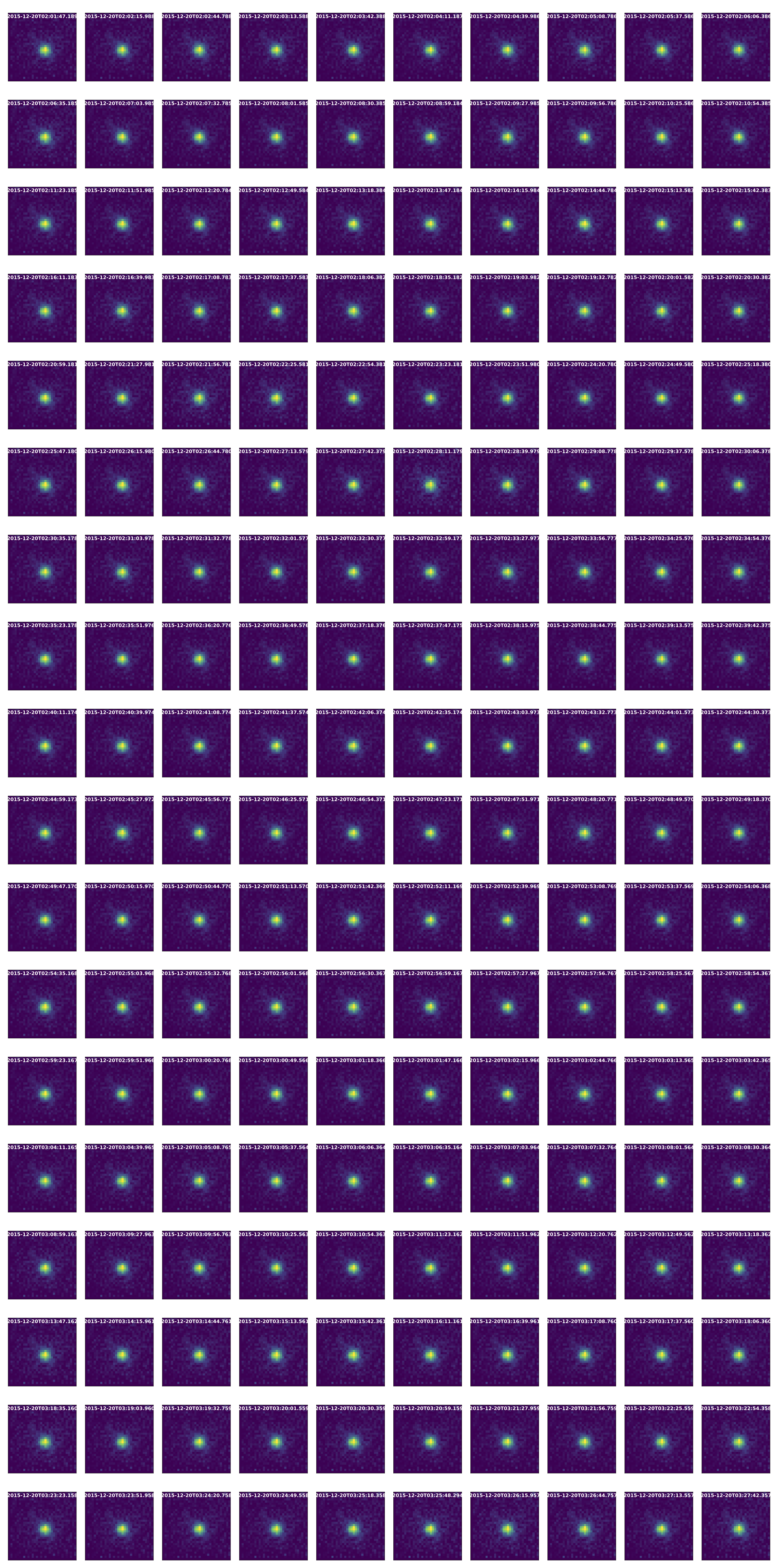}
\end{tabular}
\caption{Reconstruction of the PSF of the central star during the observations for the first (left) and second (right) epochs.}
\label{Dtts}
\end{figure*}

We show in Figure \ref{Dtts} the DTTS images retrieved for DH Tau for the four epochs obtained with SPHERE. While in the first epoch large PSF variations and significant low-wind effect can be seen, in the second epoch, the PSF as resulting from the DTTS images is remarkably stable along the full sequence of observations.

\section{LBT data reduction}

The observations at the LBT were taken on the 2019-09-19 with the L- and M-band InfraRed Camera (LMIRCam) behind the cryogrenically cooled beam combiner of the Large Binocular Telescope Interferometer \citep[LBTI,][]{Hinz}. The LBTI's primary purpose is to perform nulling interferometry \citep{Ertel,Ertel2}, but the instrument also provides the possibility to obtain among other modes sensitive, high-contrast adaptive optics imaging observations using one of the LBT apertures or both independently \citep[e.g.,][]{Stone}. Here, we used the latter mode.

The observations were taken without a coronagraph so that we could apply the \textit{multiple\_framebyframe} routines using the PSFs of the star as model. For this purpose, short (0.3 s, unsaturated frames) and long (0.9 s, saturated frames) exposure times were alternated during the 2h40min time of the observation, with a stable seeing of $\sim 0.65''$ and a FoV rotation angle of $\sim 37^{\circ}$. A total of 6620 frames were acquired, with 40 unsaturated frames every 400 saturated ones. The images of the star from the two apertures of the LBT were placed next to each other on the detector at a distance sufficient not to influence the discovery space around the DH Tau system as seen by the left and right mirrors. Nodding in the down-up direction on the detector (NOD-A and NOD-B position, respectively) was done to subtract the variable sky and telescope background and detector artifacts. During each sequence of 40 unsaturated or 400 saturated frames, the star was placed in the NOD-A position for the first half of the frames and in the NOD-B position for the second half. Due to the presence of clouds and internal errors, a total of 460 frames were discarded before starting the reduction.

These frames are $2048\times2048$ pixels, joining the images of the star, either in NOD-A or NOD-B position, gathered  from the right and left mirrors. Each frame was actually composed of two images, the first of which with a minimum integration time of 27ms, so that it could be used as a quasi-instantaneous bias. Thus, as a first step of the data reduction, we subtracted the two frames to remove bias drifts at frequencies higher than the nodding frequency. Then, we subtracted to each frame in one NOD position the mean of the closest sequence taken in the opposite NOD. For example, considering the sequence of frames from 1 to 400, we subtracted to each frame in NOD-A (from 1 to 200) the mean of the frames in NOD-B (from 201 to 400). This procedure removes all the remaining bias, the majority of detector artifacts, and the sky- and telescope background.

The LMIRCam observations are strongly affected by distortion effects, depending on the position of the star in the FoV. We then applied a distortion correction provided by the LBTI team \citep{Spalding} and available on the LBTO webpage\footnote{https://sites.google.com/a/lbto.org/lbti/data-retrieval-reduction/distortion-correction-and-astrometric-solution}.

We divided the frames into four datasets, one for each mirror (left or right) and NOD position, recentering the images on the position of the star and, subsequently, cropping them to a FoV of $7.5''\times7.5''$ ($701\times701$). In order to detect the brown dwarf in each frame where the central star was saturated, we stacked together each 20 consecutive images and gathered them into two separate cubes: the scientific one composed of long exposure time frames and the one with unsaturated frames to be used as model PSFs. We thus obtained four scientific cubes, each with 140 frames, and four model PSFs cubes, each with 14 frames.

To further improve the contrast, we also removed the bad pixels with a sigma-clipping and the "striping" that occurs in LMIRCam images due to the vertical detector channels. Fiinally, we applied the PSF subtraction using the \textit{multiple\_framebyframe} routine, considering one model PSF every ten frames of the scientific cube.

Unfortunately, the NOD-B images for both apertures were obtained with the target close to the edge of  LMIRCam’s FoV and we have discovered that uncorrected image distortion appears to affect the image quality in this case.  Thus, these frames could not be used in our analysis. The NOD-A cubes were, instead, good-quality observations and we used them both individually and jointly to determine the parameters presented in the text.

\end{appendix}

\end{document}